\documentclass[aps,twocolumn,prd,10pt,showpacs,showkeys,preprintnumbers,superscriptaddress,floatfix,longbibliography,nofootinbib,notitlepage]{revtex4-1}

\usepackage{amsmath,amsfonts,amssymb,mathrsfs,graphicx,color,longtable}
\usepackage{hyperref}
\usepackage{bm}
\usepackage{array}
\usepackage{enumitem}
\newcommand\bout{\bgroup\markoverwith{\textcolor{blue}{\rule[0.5ex]{4pt}{0.8pt}}}\ULon}

\newcommand{\beq}{\begin{equation}}
\newcommand{\eeq}{\end{equation}}
\usepackage[explicit]{titlesec}

\newcommand*{\absq}[1]{\lvert #1 \rvert^2}

\newcommand{\Unu}{U_{\rm PMNS}}

\newcolumntype{C}[1]{>{\centering\let\newline\\\arraybackslash\hspace{4pt}}m{#1}}

\graphicspath{{figures/}}

\begin{document}

\title{Leptonic Unitarity Triangles}

\author{Sebastian A. R. Ellis}
\email{sarellis@slac.stanford.edu}
\thanks{\scriptsize \!\! \href{https://orcid.org/0000-0003-3611-2437}{0000-0003-3611-2437}}
\affiliation{SLAC National Accelerator Laboratory, 2575 Sand Hill Road, Menlo Park, CA 94025, USA}

\author{Kevin J. Kelly}
\email{kkelly12@fnal.gov}
\thanks{\scriptsize \!\! \href{https://orcid.org/0000-0002-4892-2093}{0000-0002-4892-2093}}
\affiliation{Theoretical Physics Department, Fermi National Accelerator Laboratory, P. O. Box 500, Batavia, IL 60510, USA}

\author{Shirley Weishi Li}
\email{shirleyl@slac.stanford.edu}
\thanks{\scriptsize \!\! \href{https://orcid.org/0000-0002-2157-8982}{0000-0002-2157-8982}\\}
\affiliation{SLAC National Accelerator Laboratory, 2575 Sand Hill Road, Menlo Park, CA 94025, USA}

\date{December 4, 2020}

\preprint{FERMILAB-PUB-20-156-T, SLAC-PUB-17521}

\begin{abstract}
We present a comprehensive analysis of leptonic unitarity triangles, using both current neutrino oscillation data and projections of next-generation oscillation measurements. Future experiments, sensitive to the degree of CP violation in the lepton sector, will enable the construction of precise triangles. We show how unitarity violation could manifest in the triangles and discuss how they serve as unitarity tests. We also propose the use of Jarlskog factors as a complementary means of probing unitarity. This analysis highlights the importance of testing the unitarity of the leptonic mixing matrix, an understanding of which is crucial for deciphering the nature of the neutrino sector.
\end{abstract}

\maketitle

%%%%%%%%%%%%%%%%%%%%%%%%%%%%%%%%%%%%%%%%%%%%%%%%%%%%%%%%
%%%%%%%%%%%%%%%%%%%%%%%%%%%%%%%%%%%%%%%%%%%%%%%%%%%%%%%%

\section{Introduction}
\label{sec:Introduction}

The discovery of neutrino oscillations confirmed that lepton flavor and mass eigenstates are distinct.  Their mixing is canonically parameterized by the $3\times3$ unitary Pontecorvo-Maki-Nakagawa-Sakata (PMNS) matrix~\cite{Pontecorvo:1967fh,Maki:1962mu}, analogous to the Cabibbo-Kobayashi-Maskawa (CKM) matrix~\cite{Cabibbo:1963yz, Kobayashi:1973fv} for quarks. A crucial difference between the leptonic and quark sectors, however, is our knowledge of the origin of such mixing. The appearance of the known active neutrinos in $SU(2)_L$ doublets means the flavor eigenstates are known. However, the structure of the neutrino mass terms is unknown, since the Standard Model as formulated does not contain right-handed singlet fermions, and therefore does not allow for a renormalizable neutrino-Higgs Yukawa interaction. The misalignment of the flavor and mass eigenstates in the lepton sector, i.e., the origin of the PMNS matrix, therefore remains an open question. Predictions of neutrino masses, e.g., those involving right-handed neutrinos, often lead to a non-unitary $3\times3$ leptonic mixing matrix (LMM), which is a sub-matrix of a larger unitary matrix (see e.g.~\cite{Wyler:1982dd, Langacker:1988up, Hewett:1988xc, Buchmuller:1991tu,Ingelman:1993ve,Nardi:1993ag, Chang:1994hz, Tommasini:1995ii,Loinaz:2003gc}). Searches for deviation from unitarity of the LMM therefore have the potential to directly probe our fundamental understanding of neutrino masses. Throughout this work, we refer to a general $3\times3$ LMM as $U_{\rm LMM}$ and one assumed to be unitary as $U_{\rm PMNS}$.

$U_{\rm PMNS}$ is parameterized by three angles, $\theta_{12},~\theta_{13},~\theta_{23}$, and a phase $\delta_{\rm CP}$.\footnote{If neutrinos are Majorana, two additional phases, which do not affect oscillations, appear. We disregard them for the remainder of this work.} The degree to which the combination of charge C and parity P symmetry, CP, is violated is proportional to the Jarlskog invariant~\cite{Jarlskog:1985ht}, 
\begin{equation}
    J_{\rm PMNS} \equiv c_{12} c_{13}^2 c_{23} s_{12} s_{13} s_{23} \sin \delta_{\rm CP} \ ,
    \label{eq:Jpmns}
\end{equation}
where $s_{ij} = \sin\theta_{ij}$, $c_{ij} = \cos\theta_{ij}$. This quantity is relevant to understanding the baryon asymmetry of the universe: CP-violation is one of the requirements for such an asymmetry to exist~\cite{Sakharov:1967dj}. Indeed, studies of leptogenesis leading to baryogenesis have shown that the Dirac phase $\delta_{\rm CP}$ that is measured by the Jarlskog invariant can be directly involved in the generation of an asymmetry~\cite{Branco:2001pq, Branco:2002kt, Frampton:2002qc, Endoh:2002wm, Fujihara:2005pv, Abada:2006ea, Harigaya:2012bw, Barreiros:2020mnr}. Furthering our understanding of the PMNS matrix is therefore important for probing fundamental questions.

In the quark sector, many experimental tests of the unitarity of the mixing matrix 
have been 
performed~\cite{Amhis:2016xyh}, with results often visualized using unitarity triangles~\cite{Hocker:2001xe,Bona:2006ah,CKMfitter,UTfit,Tanabashi:2018oca,Wolfenstein:1983yz,Buras:1994ec,Charles:2004jd}. In these, many measurements meet at a point in the complex plane if the matrix is unitary. Areas of such triangles are proportional to the Jarlskog invariant of the CKM matrix.

In the lepton sector, neutrino oscillation experiments can provide direct tests of LMM unitarity.  Assuming unitarity, several elements of $U_{\rm PMNS}$ have been measured to $\mathcal{O}$(10\%) precision~\cite{Gando:2010aa,Adey:2018zwh,Aharmim:2011vm,Vinyoles:2016djt,Acero:2019ksn,Agafonova:2018auq,Abe:2018wpn,Abe:2019vii,Gando:2013nba,Cleveland:1998nv, Altmann:2005ix, Abdurashitov:2009tn, Bellini:2011rx,Abe:2016nxk,Aartsen:2017nmd,Aartsen:2019tjl,Jiang:2019xwn}. Next-generation experiments will attain $\mathcal{O}$(1\%) precision, and importantly, will begin to measure the degree of CP-violation~\cite{Acciarri:2015uup,Abi:2020evt,deGouvea:2019ozk,Ishihara:2019aao,Capozzi:2018dat,Abe:2018uyc,An:2015jdp}. This allows for precision tests of LMM unitarity~\cite{Qian:2013ora,Parke:2015goa}, as well as the construction of leptonic unitarity triangles~\cite{Farzan:2002ct,He:2013rba,Gonzalez-Garcia:2014bfa,He:2016dco,Esteban:2016qun,nufit}.

\begin{figure*}
  \includegraphics[width=\textwidth]{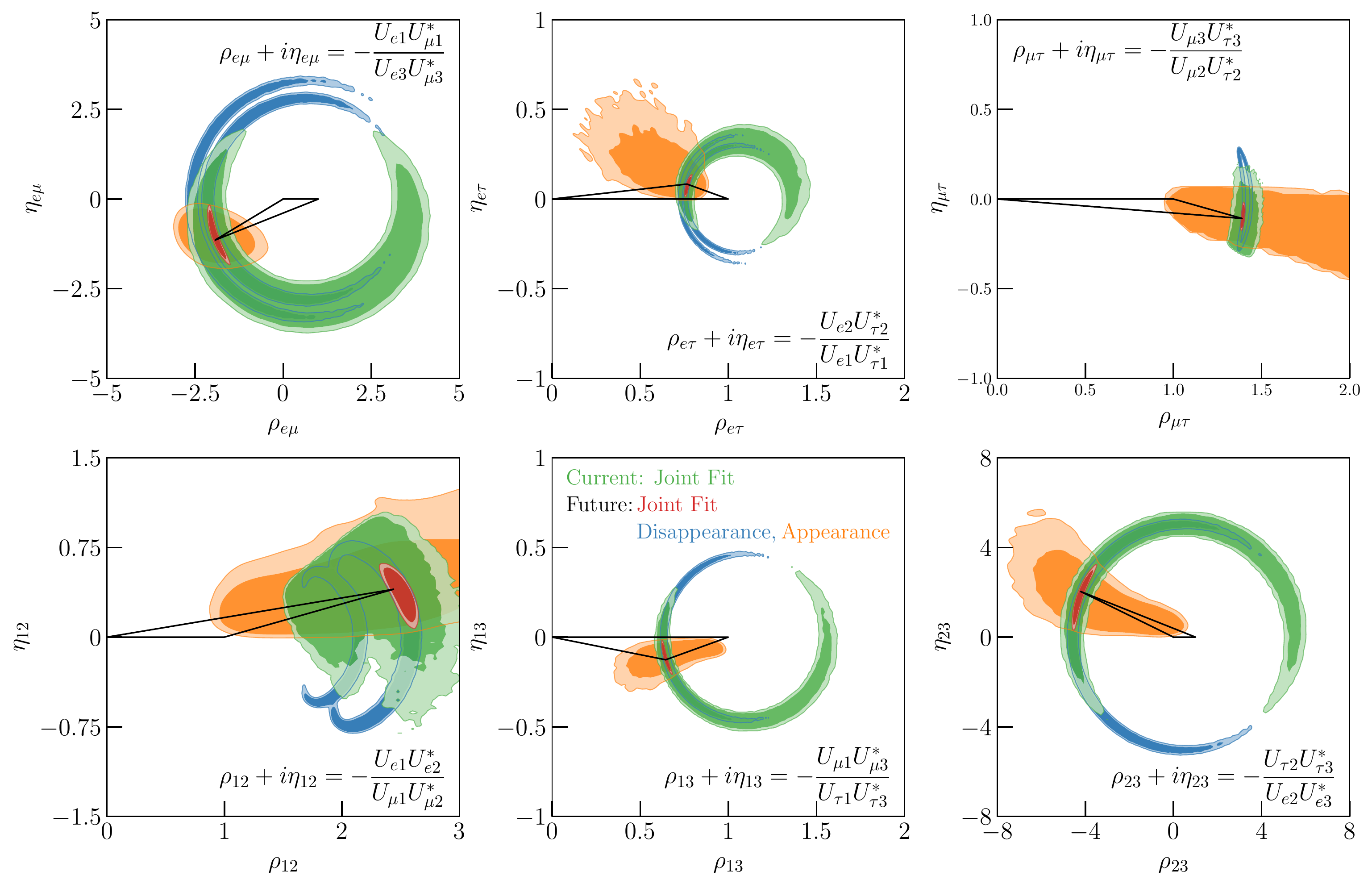}
  \caption{Current and future 95\% (dark contours) and 99\% (light) credible regions of six leptonic unitarity triangles. Green contours consist of all current data~\cite{Gando:2010aa,Adey:2018zwh,Aharmim:2011vm,Vinyoles:2016djt,Abe:2018wpn,Acero:2019ksn,Agafonova:2018auq}. In red are the analogous contours including also projections of future data from IceCube~\cite{Aartsen:2019tjl,summer_blot_2020_3959546}, JUNO~\cite{An:2015jdp}, DUNE~\cite{Acciarri:2015uup,Abi:2020evt,deGouvea:2019ozk} and T2HK~\cite{Abe:2018uyc}. Blue contours include subsets of data that measure disappearance probabilities, including reactor~\cite{Gando:2010aa,Adey:2018zwh,An:2015jdp}, solar~\cite{Aharmim:2011vm} experiments, and DUNE/T2HK $\nu_\mu \rightarrow \nu_\mu$. Orange contours include appearance measurements, e.g.., DUNE \& T2HK $\nu_\mu \rightarrow \nu_e$ and IceCube $\nu_\tau$ appearance. This demonstrates the complementarity of these measurements in constructing unitarity triangles. The black triangles point to the best-fit point on the plane. Unitarity of the LMM is assumed. A subset of triangles without unitarity assumed is shown in Fig.~\ref{fig:RhoEta_UnitarityAssumptions}.
  }
  \label{fig:UnTriangles}
\end{figure*}

In this work, we present a comprehensive analysis of leptonic unitarity triangles using current neutrino oscillation data and projections of future experiments. Our main results are shown in Fig.~\ref{fig:UnTriangles}, where we present six unitarity triangles, and in Fig.~\ref{fig:Jarlskogs}, where we show nine Jarlskog factors and the PMNS Jarlskog invariant. This set of measurements allows for a complete understanding of the LMM in a compact form. In our companion paper~\cite{Ellis:2020hus}, we discuss how well these datasets constrain all unitarity conditions of $U_{\rm LMM}$. Our results show the importance of separately analyzing appearance and disappearance data, and demonstrate the power of future oscillation measurements to constrain fundamental physics related to the neutrino sector.

%%%%%%%%%%%%%%%%%%%%%%%%%%%%%%%%%
%%%%%%%%%%%%%%%%%%%%%%%%%%%%%%%%%

%%%%%%%%%%%%%%%%%%%%%%%%%%%%%%%%%%%%%%%%%%%%%%%%%%%%%%%%
%%%%%%%%%%%%%%%%%%%%%%%%%%%%%%%%%%%%%%%%%%%%%%%%%%%%%%%%

\section{The Leptonic Mixing Matrix and Unitarity Triangles}
\label{sec:LMM}

The LMM describing the mixing of leptonic flavor and mass eigenstates,
\begin{equation}
    U_{\rm LMM} = 
    \begin{pmatrix} 
    U_{e1} & U_{e2} & U_{e3} \\
    U_{\mu1} & U_{\mu2} & U_{\mu3} \\
    U_{\tau1} & U_{\tau2} & U_{\tau3}
    \end{pmatrix} \ ,
\end{equation}
is defined by its appearance in the charged current interaction. In full generality, $U_{\rm LMM}$ may 
be a non-unitary $3\times3$ sub-matrix of a larger $M\times N$ complex matrix.
If only three generations of leptons exists, $U_{\rm LMM} \equiv U_{\rm PMNS}$.
This mixing leads to neutrino oscillation, the precise measurement of which is the focus of current and future detection efforts. The primary measurements in oscillation experiments are the appearance and disappearance probabilities, $P_{\alpha\beta}$ and $P_{\alpha\alpha}$ ($\alpha \neq \beta$; $\alpha,\beta \in e, \mu, \tau$).
Distinguishing between these is important for testing unitarity.

For a $3\times3$ unitary matrix, a set of six triangles may be defined from the conditions $U^\dagger U = U U^\dagger = \mathbb{I}$, corresponding to the closure of the products of cross-terms of the matrix:
\begin{equation}
    \begin{array}{c c}\displaystyle\sum_i U_{\alpha i}U_{\beta i}^* = 0\ ,\\ (\alpha \neq \beta, \text{rows})\end{array} \quad  \begin{array}{cc}
   \displaystyle\sum_\alpha U_{\alpha i}U_{\alpha j}^* = 0 \ . \\ (i \neq j, \text{columns}) \end{array} 
   \label{eq:closuretext}
\end{equation}
The elements $U_{\alpha i}$ are complex, so the above conditions can be shown as closed triangles in complex planes. Unitarity triangles are constructed by normalizing to one of the three terms in the sums of Eq.~(\ref{eq:closuretext}), and defining a vertex of the triangle as $\rho_{xy} + i\eta_{xy}$, $(x\neq y)$,
\begin{equation}
    \rho_{x y} + i \eta_{ x y} = 
    \begin{cases}
    &-\displaystyle\frac{U_{x i} U_{y i}^*}{U_{x j} U_{y j}^*} \ ,\quad \text{rows} \\
    \\
    &-\displaystyle\frac{U_{\alpha x} U_{\alpha y}^*}{U_{\beta x} U_{\beta y}^*}\ ,\quad \text{columns}
    \end{cases} \ ,
    \label{eq:rhoetatext}
\end{equation}
such that a closed triangle has vertices at the origin, $(1,0)$, and $(\rho_{xy},\eta_{xy})$. There is ambiguity in the choice of the denominator, and hence $(\rho_{xy},\eta_{xy})$ for a given row/column. We explain our choices below, made in an attempt to cover measurements of all parameters of $U_{\rm PMNS}$, and give the full definitions of $(\rho_{xy},\eta_{xy})$ in Appendix~\ref{sec:rhoeta}.

We define possible choices of row unitarity triangles as:
\begin{align}
\label{eq:RowTriangleShorthand}
	&(\rho_{\alpha\beta} + i \eta_{\alpha\beta})^{(1)} = T_{\alpha\beta}^{(1)} = - \frac{U_{\alpha1} U_{\beta 	1}^*}{U_{\alpha3} U_{\beta 3}^*} \ , \nonumber \\
	&(\rho_{\alpha\beta} + i \eta_{\alpha\beta})^{(2)} = T_{\alpha\beta}^{(2)} = - \frac{U_{\alpha2} U_{\beta 	2}^*}{U_{\alpha 1} U_{\beta 1}^*}, \nonumber \\
	&(\rho_{\alpha\beta} + i \eta_{\alpha\beta})^{(3)} = T_{\alpha\beta}^{(3)} = - \frac{U_{\alpha3} U_{\beta 	3}^*}{U_{\alpha 2} U_{\beta 2}^*} \ , \\
	&T_{\alpha \beta}^{(-m)} = \left( T_{\alpha \beta}^{(m)} \right)^{-1} \ ,	
\label{eq:RowTriangleShorthandRec} 
\end{align}
where $\alpha, \beta \in [e,\mu, \tau]$. The column unitarity triangles are defined as
\begin{align}
\label{eq:ColTriangleShorthand}
	&(\rho_{ij} + i \eta_{ij})^{(1)} = T_{ij}^{(1)} = - \frac{U_{\mu i} U_{\mu j}^*}{U_{e i} U_{e j}^*} \ , 		\nonumber \\ 
	&(\rho_{ij} + i \eta_{ij})^{(2)} = T_{ij}^{(2)} = - \frac{U_{\tau i} U_{\tau j}^*}{U_{\mu i} U_{\mu j}^*}, 	\nonumber \\
	&(\rho_{ij} + i \eta_{ij})^{(3)} = T_{ij}^{(3)} = - \frac{U_{e i } U_{e j}^*}{U_{\tau i} U_{\tau j}^*} \ , \\
	 &T_{i j }^{(-m)} = \left( T_{i j }^{(m)} \right)^{-1} \ ,
	 \label{eq:ColTriangleShorthandRec}
\end{align}
where $i, j \in [1,2,3]$. 

We also define general Jarlskog factors $J_{\alpha i}$ as
\begin{equation}
    \varepsilon_{\alpha\beta\gamma}\varepsilon_{ijk} J_{\alpha i} = \text{Im}\left( U_{\beta j}U_{\gamma k} U_{\beta k}^* U_{\gamma j}^*\right) \ ,
    \label{eq:Jarlskog}
\end{equation}
which are related to the areas $A_{T}$ of the possible triangles. For the row triangles as defined above, we obtain the following relations between $\text{Im}(T_{\alpha\beta}^{(m)}) = \pm 2 A_{T_{\alpha \beta}^{(m)}}$ and the Jarlskog factors: 
\begin{align}
    &\text{Im}(T_{\alpha\beta}^{(1)}) =  \frac{J_{\gamma 2}}{|U_{\alpha 3}|^2|U_{\beta 3}|^2} \ , \nonumber \\
    &\text{Im}(T_{\alpha\beta}^{(2)}) =  \frac{J_{\gamma 3}}{|U_{\alpha 1}|^2|U_{\beta 1}|^2} \ , \nonumber \\
    &\text{Im}(T_{\alpha\beta}^{(3)}) =  \frac{J_{\gamma 1}}{|U_{\alpha 2}|^2|U_{\beta 2}|^2} \ , 
\end{align}
where $\alpha \neq \beta \neq \gamma$.

Repeating the same analysis for the column triangles as for the rows, we can derive the following relations between the triangles and Jarlskog factors:
\begin{align}
    &\text{Im}(T_{ij}^{(1)}) = \frac{J_{\tau k}}{|U_{e i}|^2|U_{e j}|^2} \ , \nonumber \\
    &\text{Im}(T_{ij}^{(2)}) = \frac{J_{e k}}{|U_{\mu i}|^2|U_{\mu j}|^2} \ ,\nonumber \\
    &\text{Im}(T_{ij}^{(3)}) = \frac{J_{\mu k}}{|U_{\tau i}|^2|U_{\tau j}|^2} \ ,
\end{align}
where $i \neq j \neq k$.

The information contained in the triangles defined in Eqs.~(\ref{eq:RowTriangleShorthand}, \ref{eq:ColTriangleShorthand}), is duplicated by taking the reciprocal triangles of Eqs.~(\ref{eq:RowTriangleShorthandRec}, \ref{eq:ColTriangleShorthandRec}), so one need only consider one set. If we do not assume unitarity when performing the triangle analysis, we can see that 9 unitarity triangles must be constructed in order to fully measure all 9 Jarlskog factors once in combination with all matrix element norms twice.

If we assume unitarity when constructing triangles, we would find that all $J$-factors are identical by definition, and equal to the Jarlskog invariant in the PMNS parameterization. Thus constructing triangles allows us to measure $J_{\rm PMNS}$ and products of norms. In this case, to display all the information contained in the 18 possible triangles, we must pick 6 triangles to cover all 9 norms with minimal redundancy. Any set of 6 triangles which includes 9 separate norms actually contains 12 norms, such that there are always three norms which are measured twice. An example of a set of triangles which would encapsulate all possible information would be $T \supset T_{e\mu}^{(1)},~T_{e\mu}^{(2)},~T_{e\mu}^{(3)},~T_{e\tau}^{(1)},~T_{e\tau}^{(2)},~T_{e\tau}^{(3)}$. With this set, $|U_{ei}|$, $i=1,2,3$ would be repeated twice. A more ``flavor-democratic" approach to choosing a set of six triangles, and the one we use in our analysis, is to choose one triangle from each row and column:
\begin{equation}
    T \supset T_{e\mu}^{(1)},~ T_{e\tau}^{(2)},~T_{\mu \tau}^{(3)},~T_{12}^{(-1)},~T_{13}^{(-2)},~T_{23}^{(-3)} \ .
    \label{TriangleChoice}
\end{equation}
The full expressions for these triangles are given in Appendix~\ref{sec:rhoeta} in terms of the PMNS parameterization, and can be obtained in the LMM parameterization from Eqs.~(\ref{eq:RowTriangleShorthand}) and (\ref{eq:ColTriangleShorthandRec}).

In this way, we measure all information in the LMM matrix under the assumption of unitarity, while only repeating measurements of $|U_{e3}|,~|U_{\mu 2}|$ and $|U_{\tau 1}|$. This specific set of choices is further motivated by the discussion in Section~\ref{sec:Results} around Fig.~\ref{fig:unitarity_violation} of how to use unitarity triangles to observe unitarity violation. Given the degree of non-unitarity allowed by current measurements, which is then used to construct Fig.~\ref{fig:unitarity_violation}, it was determined that the above set of 6 triangles was most instructive for observing tension between the appearance and disappearance data in the various $(\rho,\eta)$ planes.

Other choices of triangles have been made previously in Refs.~\cite{Gonzalez-Garcia:2014bfa,Esteban:2016qun,nufit}. For the 1-3 triangle, their choice of $(\rho, \eta)$ is
\begin{equation}
    \rho_{13} + i \eta_{13} = - \frac{U_{e 1} U_{e 3}^*}{U_{\mu 1} U_{\mu 3}^*} .
\end{equation}
This choice corresponds to $T_{13}^{(-1)}$ as defined above, and is therefore measuring $J_{\tau 2}$.
Note that this definition is the leptonic equivalent of the $d-b$ CKM triangle that is commonly shown. When we adopt this definition, our joint-fit region from current experiments is consistent with the result of Refs.~\cite{Esteban:2016qun,nufit}, which can be broken down to a disappearance circle that is centered at (0, 0) and an appearance region that is more visibly radially oriented.

In order to fully characterize a potentially non-unitary LMM in a relatively economical yet intuitive fashion, it is clear that we must account for the fact that assuming unitarity sets all $J$-factors equal in the triangles. Thus we must separately measure all nine possible $J$-factors. By showing an appropriate set of 6 unitarity triangles when assuming unitarity (Fig.~\ref{fig:UnTriangles}), 9 $J$-factors computed without assuming unitarity, and $J_{\rm PMNS}$ (Fig.~\ref{fig:Jarlskogs}), we graphically represent all possible information in the LMM. There is an added benefit to computing $J$-factors separately, as they include information obtained from sterile searches that is not otherwise visible in the triangle planes.

%%%%%%%%%%%%%%%%%%%%%%%%%%%%%%%%%%%%%%%%%%%%%%%%%%%%%%%%
%%%%%%%%%%%%%%%%%%%%%%%%%%%%%%%%%%%%%%%%%%%%%%%%%%%%%%%%

\section{Data Analysis and Methodology}
\label{sec:Data}

We take global neutrino data and recast their joint measurements onto leptonic triangles. Our goal is not to do the most precise, comprehensive global fit on mixing parameters, so we interpret the majority of experimental results as rate-only measurements. Concretely, we assume a given experiment measures an oscillation probability $P_{\alpha\beta}$ at a fixed ($L$, $E_\nu$). This reasonably reproduces the reported experimental results, so we apply it to T2K~\cite{Abe:2018wpn,T2KNu2020} and NOvA~\cite{Acero:2019ksn,NOvANu2020} (which are sensitive to $\delta_{\rm{CP}}$), as well as Daya Bay~\cite{Adey:2018zwh}, solar neutrino measurements~\cite{Aharmim:2011vm,Aharmim:2011vm,Vinyoles:2016djt,SKNu2020}, and OPERA~\cite{Agafonova:2018auq}. For KamLAND, we include a more detailed analysis that utilizes the measured neutrino spectrum~\cite{Gando:2010aa}. 

We also project the inclusion of future data in our simulations, namely, The Deep Underground Neutrino Experiment (DUNE) $\nu_\mu$-disappearance and $\nu_e$-appearance channels~\cite{Acciarri:2015uup,Abi:2020evt,DUNEfluxes,Formaggio:2013kya,Berryman:2015nua,deGouvea:2015ndi,deGouvea:2016pom}, $\nu_\tau$-appearance channel~\cite{deGouvea:2019ozk}, The Jiangmen Underground Neutrino Observatory (JUNO)~\cite{An:2015jdp,Capozzi:2013psa,An:2013uza,Huber:2011wv,Mueller:2011nm,Strumia:2003zx}, Tokai-to-Hyper-Kamiokande (T2HK)~\cite{Abe:2018uyc}, and the IceCube Upgrade's capabilities for measuring $\nu_\tau$ appearance~\cite{Aartsen:2019tjl,summer_blot_2020_3959546}. The following subsections detail the current and future data included. We direct the reader to Ref.~\cite{Ellis:2020hus}, where the current and future data included are identical to those here, and more details are provided.

Using a given combination of data sets, we construct a likelihood function depending on a set of oscillation parameters. Our fits in Figs.~\ref{fig:UnTriangles}-\ref{fig:unitarity_violation} depend on six parameters: $\sin^2\theta_{12}$,\ $\sin^2\theta_{13},$\ $\sin^2\theta_{23},$\ $\delta_{\rm{CP}},$\ $\Delta m_{21}^2,$\ $\Delta m_{31}^2$. We include Gaussian priors on the mass-squared splittings from respective experimental results when analyzing current data, and use the Bayesian inference tool {\sc pyMultinest}~\cite{Feroz:2008xx,Buchner:2014nha} to construct credible regions in this parameter space. The posterior distributions are then projected onto ($\rho_{xy}$, $\eta_{xy}$).

Analyzing current data, the maximum-likelihood parameters are $\sin^2\theta_{12} = 0.308$, $\sin^2\theta_{13} = 0.0219$,
$\sin^2\theta_{23} = 0.551$, $\delta_{\rm CP} = 200^\circ$, consistent\footnote{Previous versions of our analysis preferred a closer-to-maximal value of $\delta_{\rm CP}$, in line with published global fits prior to the latest release of data from T2K and NOvA~\cite{Esteban:2018azc,deSalas:2017kay,Capozzi:2020qhw}.} with global fits~\cite{Esteban:2020cvm}. We assume these to be the true parameters when simulating future experiments.

\subsection{Current Experimental Results Included}
\label{app:CurrentData}

Here we list the current experimental results that we include in our data analysis, and specify to which parameters each experiment is most sensitive. We also show the results of our data analysis (when unitarity is assumed) of all of the current data included in measuring the combination of the parameters $\delta_{\rm CP}$ and $\sin^2\theta_{23}$, validating our approach.

\textbf{Mass-Squared Splittings} When mentioned in the following list, we allow the mass-squared splittings $\Delta m_{21}^2$ and $\Delta m_{31}^2$ to vary independently, allowing the possibility of both mass orderings. Based on our methods of incorporating existing measurements, KamLAND, Daya Bay, T2K, NOvA, and OPERA are sensitive to mass-squared splittings. For each experiment, we include a Gaussian prior on the relevant mass-squared splitting from the experimentally reported 1$\sigma$ range. The resulting fit region of the two mass-squared-splittings that we obtain is consistent with those from more sophisticated global fits~\cite{deSalas:2017kay,Esteban:2018azc,Capozzi:2020qhw,Esteban:2020cvm}. While tensions exist\footnote{However, these tensions are smaller with the latest analysis from Super-Kamiokande and SNO~\cite{SKNu2020}, which measure $\Delta m_{21}^2 = \left(6.11 \pm 1.21\right) \times 10^{-5}$ eV$^2$. We include this as a prior in our analysis, even though the KamLAND measurement is significantly more powerful.} between different measurements of $\Delta m_{21}^2$, we find that the analyses we perform do not change whether we also include a measurement of $\Delta m_{21}^2$ from solar neutrino measurements or not~\cite{Abe:2016nxk,SKNu2020}.

\textbf{Normalization} Many of the current (and future) experiments we consider infer a neutrino oscillation probability by measuring a far-detector-to-near-detector ratio, i.e., they measure the neutrino flux times cross sections of one flavor $\alpha$ at the near detector and one flavor $\beta$ at the far detector. The ratio of these two, up to cross section and flux effects, gives the oscillation probability $P_{\alpha\beta}$. If the LMM is not unitary, however, this is not completely true -- zero distance effects lead to $P_{\alpha\alpha}$ not being $1$ at the near detector, but 
\begin{equation}\label{eq:Norm}
    P(\nu_\alpha \to \nu_\alpha; L = 0) = \left( \sum_{i=1}^3 \left\lvert U_{\alpha i}\right\rvert^2 \right)^2.
\end{equation}
When performing an analysis that does not assume unitarity, like those surrounding Figs.~\ref{fig:Jarlskogs} and~\ref{fig:RhoEta_UnitarityAssumptions}, the inferred oscillation probability of an experiment with a near detector must be normalized by the factor in Eq.~(\ref{eq:Norm}). This normalization factor has the same impact on an analysis as including an overall systematic normalization uncertainty in an analysis, so as long as the normalization uncertainty of a given experiment is larger than the uncertainty on the quantity in Eq.~(\ref{eq:Norm}), these effects are unimportant. Normalization effects are described in much more detail in Ref.~\cite{Ellis:2020hus}.

\textbf{KamLAND} The reactor antineutrino experiment KamLAND measures the oscillation probability $P(\overline{\nu}_e \to \overline{\nu}_e)$ over a wide range of baseline length $L$ divided by neutrino energy $E_\nu$. Appendix B of Ref.~\cite{Gando:2010aa} provides measurements of this oscillation probability for different values of $L/E_\nu$. We use these measurements, which take into account matter effects, as well as either the standard, three-neutrino oscillation probability or a modified one to account for non-unitary mixing, to place constraints on mostly $\sin^2\theta_{12}$. If the matrix is not unitary, KamLAND is mostly sensitive to the product $|U_{e1}|^2 |U_{e2}|^2$. A more recent analysis is Ref.~\cite{Gando:2013nba}, but it does not contain enough information for us to reasonably reproduce its results using this approach.

\textbf{Daya Bay} For the Daya Bay experiment, we include the most recent measurement of $\sin^2\left(2\theta_{13}\right) = 0.0856 \pm 0.0029$ as a Gaussian prior in our analysis~\cite{Adey:2018zwh}. If the LMM is not unitary, the oscillation probability at Daya Bay's far detector is sensitive to the combination $4|U_{e3}|^2 (|U_{e1}|^2 + |U_{e2}|^2)$. Since Daya Bay uses a near and far detector, its measurement of $\sin^2\left(2\theta_{13}\right)$ is actually a measurement of $4|U_{e3}|^2 (|U_{e1}|^2 + |U_{e2}|^2)/(|U_{e1}|^2 + |U_{e2}|^2 + |U_{e3}|^2)^2$, as discussed above.

\textbf{Solar Neutrinos} The only solar neutrino experiments we include results from are the Sudbury Neutrino Experiment (SNO) and Super-Kamiokande (Super-K). The SNO two-flavor analysis ($\sin^2\theta_{13} \to 0$) yields $\tan^2\theta_{12}= 0.427^{+0.033}_{-0.029}$~\cite{Aharmim:2011vm}.
More generically, we include the most up-to-date measurement of the solar charged-current channel from a combined SNO and Super-K analysis, which reports $|U_{e2}|^2 (|U_{e1}|^2 + |U_{e2}|^2) + |U_{e3}|^4 = 0.2932 \pm 0.0134$~\cite{SKNu2020}.

SNO is also sensitive to neutral current scattering, which observes the effective oscillation probability $P_{\rm{NC}} = \sum_i |U_{ei}|^2_{\rm{prod.}} |U_{ei}|^2_{\rm{det.}}$. To leading order (see Ref.~\cite{Ellis:2020hus} for more detail), this becomes
\begin{align}
    P_\mathrm{NC} = &\left( |U_{e1}|^2 + |U_{e2}|^2\right)\left(|U_{e2}|^2 + |U_{\mu 2}|^2 + |U_{\tau 2}|^2\right)^2 \nonumber \\
    &+ |U_{e3}|^2 \left( |U_{e3}|^2 + |U_{\mu 3}|^2 + |U_{\tau 3}|^2\right)^2.
\end{align}
This measurement is limited by theoretical uncertainties associated with the Standard Solar Model~\cite{Vinyoles:2016djt}, so we conservatively assume that SNO measures it at the 25\% level.

\textbf{T2K} For the electron-neutrino appearance channels $P(\nu_\mu \to \nu_e)$ and $P(\overline{\nu}_\mu \to \overline{\nu}_e)$ measured at T2K, we assume that the experiment measures this probability for a fixed energy $E_\mathrm{T2K} = 600$ MeV (the mean energy of the J-PARC beam) at a distance of $L = 295$ km. We also assume a constant matter density of $\rho = 2.6$~g/cm$^3$ along the path of propagation. While this approach is an oversimplification and does not include systematic uncertainties from T2K, we find that it reproduces the results of Refs.~\cite{Abe:2018wpn,Abe:2019vii,T2KNu2020}  well. We use the predicted signal and background rates for the $\nu_e$ and $\overline{\nu}_e$ appearance presented in Ref.~\cite{T2KNu2020}, the most up-to-date results available. T2K also measures $\nu_\mu$ and $\overline{\nu}_\mu$ disappearance. We interpret this measurement as information on the quantity $4|U_{\mu 3}|^2 (|U_{\mu 1}|^2 + |U_{\mu 2}|^2)$ to agree with the results of Ref.~\cite{Abe:2018wpn}. Matter effects are much smaller in this channel, so we ignore them. If $U_{\rm LMM}$ is unitary, this translates effectively into a measurement of $\sin^2\left(2\theta_{23}\right)$. We assume T2K measures $4|U_{\mu 3}|^2 (|U_{\mu 1}|^2 + |U_{\mu 2}|^2) = 1.00 \pm 0.03$. We find that including disappearance information in this way reproduces the measurement capability of the experiment from Refs.~\cite{Abe:2018wpn,Abe:2019vii,T2KNu2020} better than assuming a fixed-length, fixed-energy measurement in this channel. See Ref.~\cite{Ellis:2020hus} for more details of this analysis, as well as validation of this approach for T2K.

\textbf{NOvA} Our methodology for NOvA is very similar to that of T2K: we assume a fixed energy for the electron-neutrino appearance measurements of $E_\mathrm{NOvA} = 1.9$ GeV and $L = 810$ km (as well as a constant matter density of $2.84$ g/cm$^3$). Expected signal event rates from Ref.~\cite{NOvANu2020} allow us to construct a log-likelihood as we vary oscillation parameters. Like with T2K, we allow $\Delta m_{31}^2$ to vary within its prior for the NOvA measurements. Since NOvA prefers a value of $\sin^2\theta_{23}$ slightly away from maximal, we include its disappearance channel measurements in our fit by assuming it measures $4|U_{\mu 3}|^2 (|U_{\mu 1}|^2 + |U_{\mu 2}|^2) = 0.99 \pm 0.02$. We find good agreement between our simplified analysis and the results of Ref.~\cite{Acero:2019ksn,NOvANu2020} for all different oscillation parameters of interest. Again, Ref.~\cite{Ellis:2020hus} provides details and validation of our NOvA analysis.

\textbf{OPERA} We include the OPERA collaboration's measurement of tau neutrino appearance via $P(\nu_\mu \to \nu_\tau)$, where 10 $\nu_\tau$ signal events were observed with an expected background of $2.0 \pm 0.4$ events. Assuming $\sin^2\left(2\theta_{23}\right) = 1$ and $\Delta m_{32}^2 = 2.5 \times 10^{-3}$~eV$^2$, OPERA expected $6.8\pm 1.4$ signal events. We include this information, assuming a mono-energetic measurement at $E_\mathrm{OPERA} = 17$ GeV and $L = 730$ km, giving results consistent with those from OPERA~\cite{Agafonova:2018auq}. Matter effects are included, even though they are small for $\nu_\tau$ appearance oscillation probabilities.

\textbf{Sterile Neutrino Searches} When unitarity is not assumed, sterile neutrino searches provide additional constraints on the unitarity of the LMM. Specifically, we include results from sterile neutrino searches at experiments in regions where such new oscillations would have ``averaged out'' in the experiment's detector. This corresponds to the high $\Delta m_{41}^2$ mass-squared splitting region of experimental sensitivities/exclusions and utilizes the zero-distance effects. Searches for anomalous \textit{appearance} of a different neutrino flavor constrain triangle closure information, where searches for anomalous \textit{disappearance} constrain row normalizations. References~\cite{Parke:2015goa,Ellis:2020hus} provide further explanation of these effects. We do not include sterile search constraints where mass-squared-splitting information is utilized because they do not apply to a generic unitarity violation scenario where we are agnostic about the mass scale of the violation.

\textit{Neutrino Appearance Searches:} Experiments that search for anomalous appearance (such as NOMAD searching for anomalous $\nu_\mu \to \nu_e$ or $\nu_\mu \to \nu_\tau$) are sensitive to these zero-distance effects, which, if the LMM is not unitarity, correspond to the non-closure of a unitarity triangle -- $\sum_i U_{\alpha i} U_{\beta i}^* \neq 0$. We include results from KARMEN~\cite{Armbruster:2002mp}, NOMAD~\cite{Astier:2003gs,Astier:2001yj}, and CHORUS~\cite{Eskut:2007rn} in these analyses. The LSND~\cite{Athanassopoulos:1996wc,Athanassopoulos:1997pv} and MiniBooNE~\cite{Aguilar-Arevalo:2018gpe} experiments have famously observed an excess of electron-like events in a $\nu_\mu$ beam, which can be interpreted as observing a non-closure between the electron and muon rows of $\left\lvert \sum_i U_{ei} U_{\mu i}^*\right\rvert^2 \approx 2.6 \times 10^{-3}$~\cite{Aguilar-Arevalo:2018gpe}. We do not include information from MiniBooNE and LSND in our analyses due to the tension between these appearance searches and disappearance searches -- Ref.~\cite{Ellis:2020hus} explores this in much more detail.

\textit{Neutrino Disappearance Searches:} For muon neutrino disappearance, we include results from the MINOS/MINOS+ experiments~\cite{Adamson:2017uda} that constrain the muon row normalization. In addition, hints for the existence of a sterile neutrino with a new mass-squared splitting around $\Delta m_{41}^2 \approx 1$ eV$^2$ have been observed in a variety of reactor antineutrino experiments (see Refs.~\cite{Dentler:2018sju,Diaz:2019fwt,Boser:2019rta,Berryman:2019hme} for reviews of these), which could point to the electron row being not properly normalized, i.e., $\sum_i \left\lvert U_{ei}\right\rvert^2 \neq 1$. However, in order to interpret these results in terms of a constraint on unitarity, we must go to the averaged-out regime of these experimental sensitivities, which is limited by the predicted reactor antineutrino fluxes~\cite{Mueller:2011nm,Huber:2011wv}. We therefore do not include these results due to the uncertainty regarding reactor antineutrino flux predictions.

\begin{figure}[ht]
    \centering
    \includegraphics[width=0.8\columnwidth]{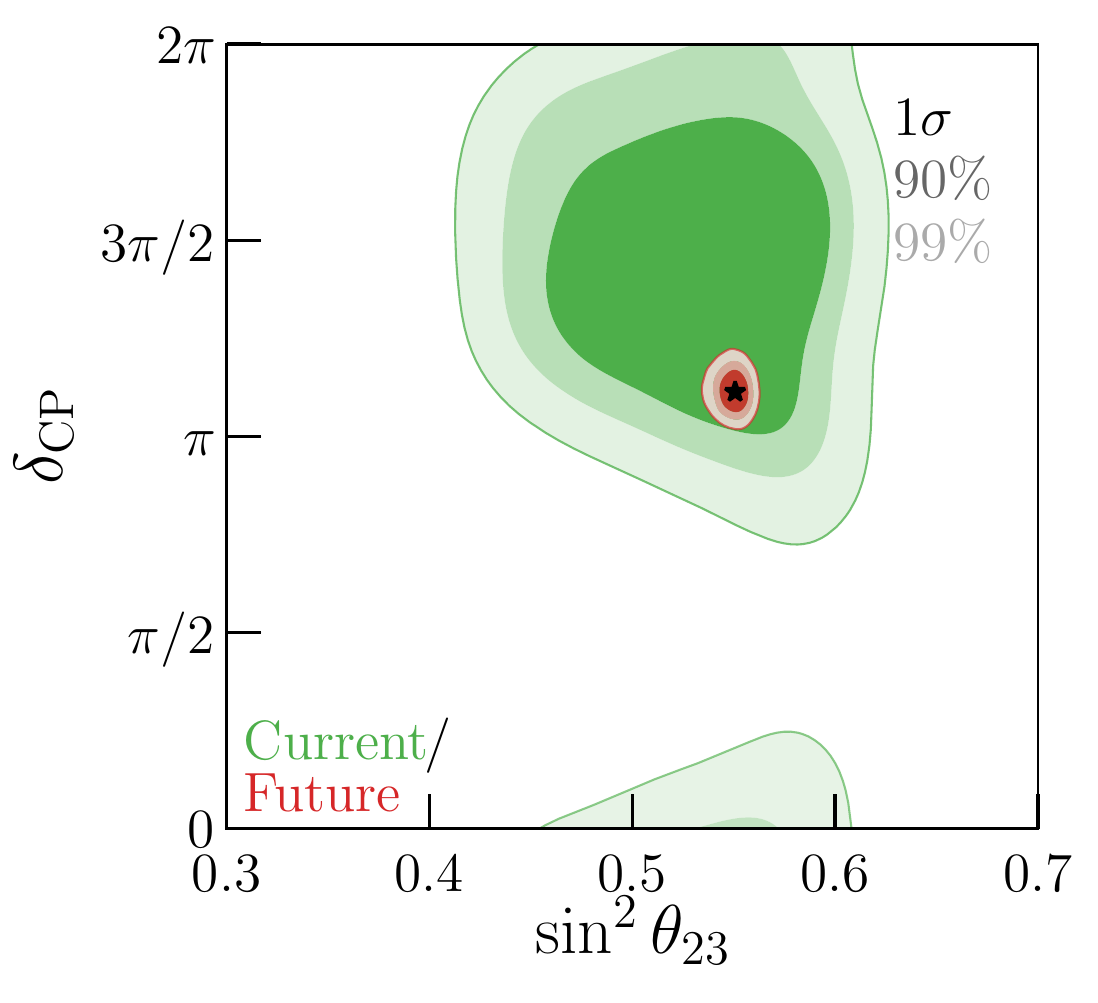}
    \caption{Results of our fit to all current data (green) discussed in Section~\ref{app:CurrentData} and including future projections (red) discussed in Section~\ref{app:FutureExps} as measurements of ($\sin^2\theta_{23}$, $\delta_{\rm CP}$) at 1$\sigma$, 90\%, and 99\%. The star indicates the best-fit point of our analysis, as well as the assumed-true combination of parameters for our future projections.}
    \label{fig:DCPS2T23}
\end{figure}

\textbf{Fit Results:} To demonstrate the validity of our methods, Fig.~\ref{fig:DCPS2T23} (green) displays the result of our fit (when unitarity is assumed) to all of the current data discussed above. We show the fit as a measurement of the parameters $\delta_{\rm CP}$ and $\sin^2\theta_{23}$, where the other, unseen parameters have been marginalized. We find that our results are consistent with more sophisticated global fits~\cite{Esteban:2018azc,deSalas:2017kay,Capozzi:2020qhw}. Specifically, we have compared our result against the most recent fit from Ref.~\cite{Esteban:2020cvm} and find that our $1\sigma$ and $90\%$ regions are slightly more conservative than theirs (likely due to the fact that their fits include \textit{all} current data where ours contain a subset) and our $99\%$ regions match very well. Our best-fit point in this parameter space, $(0.55, 3.49)$ is also very close to that of Ref.~\cite{Esteban:2020cvm}, $(0.57, 3.40)$. 

%%%%%%%%%%%%%%%%%%%%%%%%%%%%%%%%%
%%%%%%%%%%%%%%%%%%%%%%%%%%%%%%%%

%%%%%%%%%%%%%%%%%%%%%%%%%%%%%%%%%%%%%%%%%%%%%%%%%%%%%%%%
%%%%%%%%%%%%%%%%%%%%%%%%%%%%%%%%%%%%%%%%%%%%%%%%%%%%%%%%

\subsection{Future Experiment Simulations}
\label{app:FutureExps}

In our analyses, we include simulations of the DUNE, T2HK, IceCube, and JUNO experiments. In this subsection, we briefly describe how these simulations are included and some of their details. Figure~\ref{fig:DCPS2T23} includes our future projections in measuring the combination of parameters $\sin^2\theta_{23}$ and $\delta_{\rm CP}$ in red.

\textbf{DUNE} DUNE will utilize a wide-energy $\nu_\mu$ beam, with energies between roughly $0.5$ and $10$ GeV, with a baseline distance of $1300$ km. This will allow DUNE to study both $\nu_\mu$ disappearance and $\nu_e$ appearance, and allow for a powerful measurement of $\delta_{\rm CP}$. Reference~\cite{deGouvea:2019ozk} demonstrated DUNE's ability to use its beam to study $\nu_\tau$ appearance as well. We include all three of these channels in our simulations, assuming seven years of data collection (equal operation in neutrino and antineutrino modes) with a 1.2 MW beam and 40 kt of far detector fiducial mass.

Our simulation of both $\nu_\mu$ disappearance and $\nu_e$ appearance channels follows those developed for Refs.~\cite{Berryman:2015nua,deGouvea:2015ndi,deGouvea:2016pom}, and the analyses are designed to match the official collaboration sensitivities and expected signal and background event yields~\cite{Acciarri:2015uup,Abi:2020evt}. We take the neutrino fluxes from Ref.~\cite{DUNEfluxes} and neutrino cross sections from Ref.~\cite{Formaggio:2013kya}. We include energy uncertainty by using migration matrices, assuming that the energy resolution is $\sigma_E = 7\%\times E_\nu {(\rm{GeV})} + 3.5\%\sqrt{E_\nu (\rm{GeV})}$, consistent with the analyses performed in Refs.~\cite{Acciarri:2015uup,Abi:2020evt}. We include all of the different background channels discussed in Refs.~\cite{Acciarri:2015uup,Abi:2020evt} for the $\nu_\mu$ disappearance and $\nu_e$ appearance channels -- the largest of which are due to neutrino neutral-current scattering and beam contamination of opposite sign or different flavor neutrinos. Efficiencies for both signal identification and background rejection are both taken to be constant as a function of neutrino energy, where we normalize our expected signal and background event rates to those from Ref.~\cite{Acciarri:2015uup}.

Our simulation for $\nu_\tau$ appearance channel follows Ref.~\cite{deGouvea:2019ozk}. For a given true neutrino energy $E_\nu^\mathrm{true}$, we assume that the reconstructed energy follows a Gaussian distribution with a mean energy $bE_\nu^\mathrm{true}$ and an uncertainty $\sigma_E = rE_\nu^\mathrm{true}$, where $b = 45\%$ and $r = 25\%$~\cite{deGouvea:2019ozk}. We include a 25\% systematic normalization uncertainty to account for uncertainties associated with the $\nu_\tau$ charged-current cross section. We also assume a 30\% signal identification efficiency for all hadronically-decaying $\tau^\pm$ events, and that $0.5\%$ of neutral current events will contribute to backgrounds in this search.

For all channels, we include a correlated systematic normalization uncertainty on the muon neutrino flux for each beam mode (separate nuisance parameters for neutrino and antineutrino modes) of 5\%. As discussed in Section~\ref{app:CurrentData}, if we do not assume the unitarity of the LMM and an experiment has a near detector, the inferred measurement of an oscillation probability must be normalized by the appropriate channel, i.e., the normalization of the muon row of the LMM in this case. DUNE is one such experiment, however, since we include a 5\% normalization uncertainty on the muon neutrino flux, this effect is negligible in our analyses. This is because the MINOS/MINOS+ sterile neutrino search constrains this normalization effect to the $2.5\%$ level, so the systematic uncertainty of 5\% covers any impact of this type.

\textbf{T2HK} T2HK is the upcoming successor to T2K, which will include an upgraded beam and a larger water \v{C}erenkov detector~\cite{Abe:2018uyc}. Like with T2K and NOvA, T2HK and DUNE will operate in similar ranges of $L/E_\nu$ but very different ranges of $L$ and $E_\nu$. This results in matter effects being much more relevant for DUNE than T2HK, although they are not negligible and we therefore include them in our calculations.

We include simulations of both $\nu_e$ appearance and $\nu_\mu$ disappearance for both neutrino and antineutrino modes (we assume operation in a $1:3$ ratio between $\nu:\overline{\nu}$ modes, as expected by the T2HK collaboration). We assume seven years of data collection to be more consistent with our projections for DUNE. References~\cite{Kelly:2017kch,Ellis:2020hus} provide further details on this simulation.

\textbf{IceCube Upgrade} The IceCube experiment has performed a number of measurements of oscillation parameters using its atmospheric neutrino sample~\cite{Aartsen:2017nmd,Aartsen:2019tjl}, including a measurement of $\nu_\tau$ appearance that is comparable in strength to the leading measurement from Super-K, and even stronger (in terms of measuring the appearance normalization) as OPERA. Soon, IceCube expects to be able to measure this appearance at the $10\%$ level, and with the IceCube Upgrade, such a $10\%$ measurement should be quickly attainable. We include this measurement in our future projections, where IceCube is sensitive to the combination $4\absq{U_{\mu3}}\absq{U_{\tau3}}/\left( \absq{U_{\mu1}} + \absq{U_{\mu2}} + \absq{U_{\mu3}}\right)^2$.

\textbf{JUNO} JUNO will measure the oscillation of reactor $\overline{\nu}_e$ of 2--8~MeV at a propagation distance of 53~km. Matter effects can cause $\mathcal{O}(1\%)$ level modifications to this oscillation probability for these energies and propagation distance. However, they do not impact measurement sensitivity of the parameters of interest, so we do not include them. This enables a measurement of the neutrino mass ordering, the primary science goal for JUNO, as well as of $\theta_{12}$ and $\Delta m_{21}^2$.

Our analyses are designed to match the official collaboration sensitivities on $\sin^2\theta_{12}$~\cite{An:2015jdp}. For reactor flux calculation, we follow the strategy in Ref.~\cite{Capozzi:2013psa}, taking the fission isotope fractions from Ref.~\cite{An:2013uza}, $^{235}$U, $^{239}$Pu, and $^{241}$Pu spectra from Ref.~\cite{Huber:2011wv}, and $^{238}$U spectrum from Ref.~\cite{Mueller:2011nm}, and this leads to the following total reactor neutrino flux:
\begin{align}
    \Phi(E_{\overline{\nu}_e}) = 0.60\exp(&4.367-4.577E_\nu+2.1E_\nu^2-0.5294E_\nu^3 \nonumber \\ &+0.06186E_\nu^4-0.002777E_\nu^5) \nonumber\\ 
    + 0.27\exp(&4.757-5.392E_\nu+2.563E_\nu^2-0.6596E_\nu^3 \nonumber \\
    &+0.0782E_\nu^4-0.003536E_\nu^5)  \nonumber\\
    + 0.07\exp(&2.611-2.284E_\nu+0.9692E_\nu^2-0.23679E_\nu^3 \nonumber\\
    &+0.025E_\nu^4-0.001E_\nu^5) \nonumber\\
    + 0.06\exp(&2.99-2.882E_\nu+1.278E_\nu^2-0.3343E_\nu^3 \nonumber \\
    &+0.03905E_\nu^4-0.001754E_\nu^5) \ .
\end{align}

We adopt the inverse beta decay cross sections from Ref.~\cite{Strumia:2003zx}:
\begin{align}
    &\sigma = 10^{-43}\mathrm{cm}^2 p_e E_e E_\nu^{-0.07056+0.02018\log E_\nu-0.001953 \log ^3E_\nu} , \nonumber \\
    &E_e = E_\nu - 1.293~\mathrm{MeV} \ .
\end{align}
For each event, the detected energy, which is smeared with an energy resolution of  3\%$\sqrt{E (\rm{MeV})}$~\cite{An:2015jdp}, is the total energy of the positron plus its rest mass. We do not consider detector efficiencies and backgrounds, and only match the total sample size to the CDR nominal choice, 120k events (six years). For systematics, we include a correlated flux uncertainty of 2\%, an uncorrelated flux uncertainty of 0.8\%, the spectrum shape uncertainty of 1\%, and the energy scale uncertainty of 1\%~\cite{An:2015jdp}.

Similarly to the case for DUNE, the oscillation probability that JUNO measures depends on what one assumes of a near detector. Because electron row is well constrained, this assumption has a negligible impact on our results, so we perform our analysis assuming there will not be a near detector.

%%%%%%%%%%%%%%%%%%%%%%%%%%%%%%%%%%%%%%%%%%%%%%%%%%%%%%%%
%%%%%%%%%%%%%%%%%%%%%%%%%%%%%%%%%%%%%%%%%%%%%%%%%%%%%%%%

\section{Results}
\label{sec:Results}

%%%%%%%%%%%%%%%%%%%%%%%%%%%%%%%%%
\begin{figure*}
\begin{center}
\includegraphics[width=0.8\textwidth]{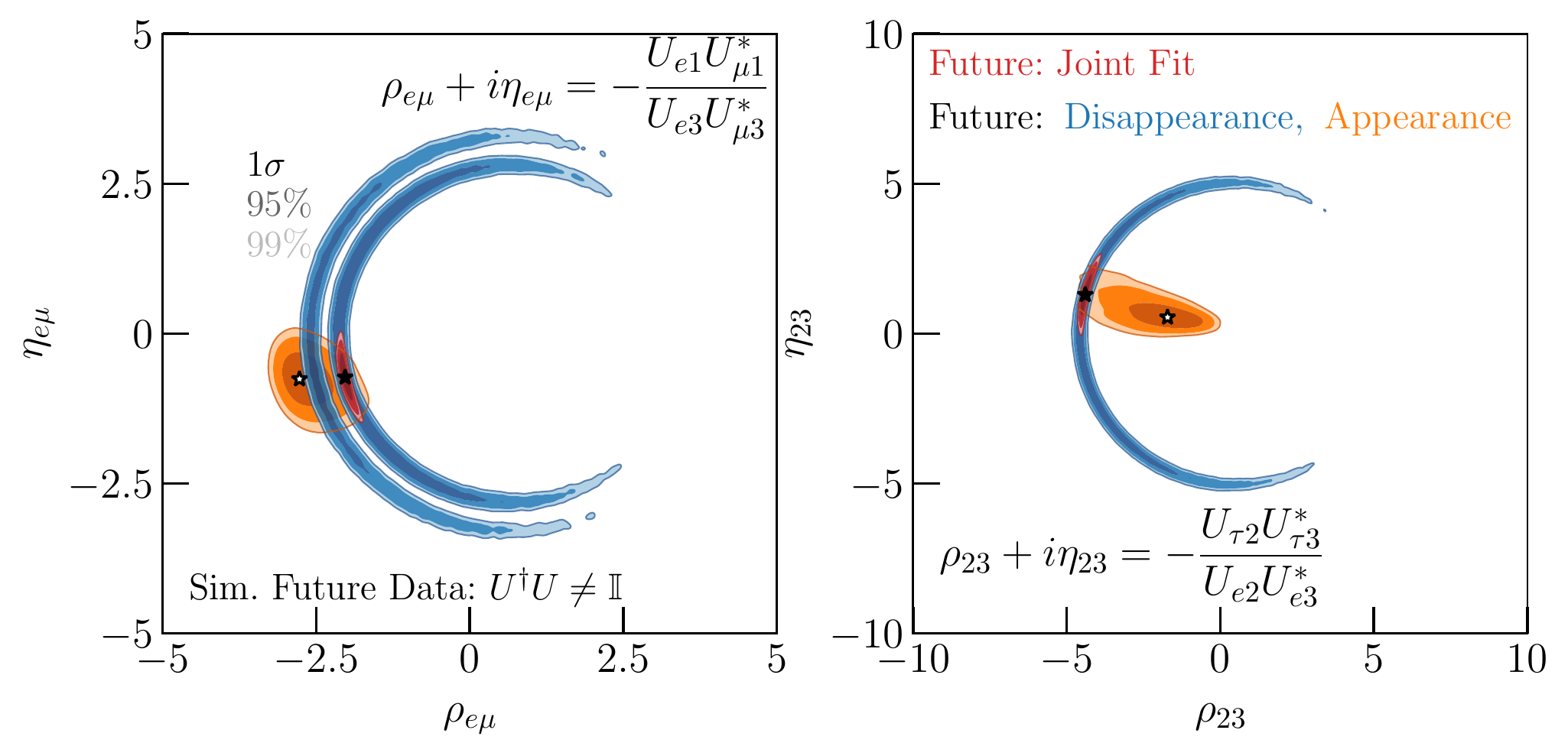}
\caption{Projected 1$\sigma$, 95\%, and 99\% regions with future measurements of $(\rho_{e\mu}, \eta_{e\mu})$ (left) and ($\rho_{_{23}}, \eta_{_{23}})$ (right). Similarly to Fig.~\ref{fig:UnTriangles}, the blue contours use disappearance analyses and the orange appearance. The filled-in (open) star indicates the best-fit point from the joint (appearance only) fit. Data are simulated with a non-unitary LMM, but analyzed assuming it is.
\label{fig:unitarity_violation}}
\end{center}
\end{figure*}
%%%%%%%%%%%%%%%%%%%%%%%%%%%%%%%%%

Figure~\ref{fig:UnTriangles} shows the 95\% and 99\% credible regions of the six unitarity triangles with all current data (green contours), and with the addition of future data (red). These results assume unitarity of the LMM and therefore the PMNS parameterization holds. We discuss the implications of assuming unitarity, and how the results would differ without this assumption in Appendix~\ref{sec:nonUparameterization}.

Figure~\ref{fig:UnTriangles} also shows the projections of future sensitivity to neutrino disappearance ($P_{\alpha \alpha}$, blue contours) and appearance ($P_{\alpha \beta}$, orange) probabilities separately. This distinction between the two measurements is crucial, both for understanding how the best fit regions arise and for determining how non-unitarity manifests itself.

The intuitively tractable $e$-$\mu$ triangle (Fig.~\ref{fig:UnTriangles} top-left) can be expressed in terms of PMNS parameters as,
\begin{align}
\rho_{e\mu} &= c_{12}^2 + \cos\delta_{\rm CP} \left( \frac{s_{12} c_{12} c_{23}}{s_{13} s_{23}}\right), \nonumber \\
\eta_{e\mu} &= \sin\delta_{\rm CP} \left( \frac{s_{12} c_{12} c_{23}}{s_{13} s_{23}}\right). \label{eq:rhoeta_emu}
\end{align}
Precise measurements of the disappearance channels $P_{ee}$ and $P_{\mu\mu}$ allow for the determination of $\theta_{12}$, $\theta_{13}$, and $\theta_{23}$, which determine a circle (we discuss why the blue circles do not close in the following paragraph). Long-baseline disappearance measurements ($P_{\mu\mu}$) are only sensitive to $\sin^2(2\theta_{23})$, leading to an octant degeneracy. This produces ambiguity in the radii of the circles in the $e$-$\mu$ and $e$-$\tau$ planes, and results in the structure in the 1-2 plane. Long-baseline $\nu_e$-appearance measurements can determine the value of $\delta_{\rm CP}$, and therefore a preferred direction in the $(\rho,\eta)$ plane. The definitions of $(\rho,\eta)$ affect the appearance of these triangles, but the general features of disappearance measurements giving ring-type structures and appearance selecting out a direction are universal (For comparisons to other choices, see Refs.~\cite{Gonzalez-Garcia:2014bfa,Esteban:2016qun,nufit} and our discussion above). $\nu_\tau$-appearance measurements, present or projected, are insufficiently precise to provide further discriminatory power. In addition, no planned measurement of $P_{\mu\tau}$ is actually sensitive to $\delta_{\rm CP}$, even with improved precision~\cite{deGouvea:2019ozk,Ishihara:2019aao}. Note that the power of $\tau$-appearance measurement manifests clearly when unitarity is not assumed; see Ref.~\cite{Ellis:2020hus} for details. In contrast with the CKM fits, where multiple observables independently measure the CP-violating phase, in the neutrino sector there is only one present or near-future observable, $P_{\mu e}$, sensitive to $\delta_{\rm CP}$.

%%%%%%%%%%%%%%%%%%%%%%%%%%%%%%%%%
\begin{figure*}[th]
\begin{center}
\includegraphics[width=0.6\textwidth]{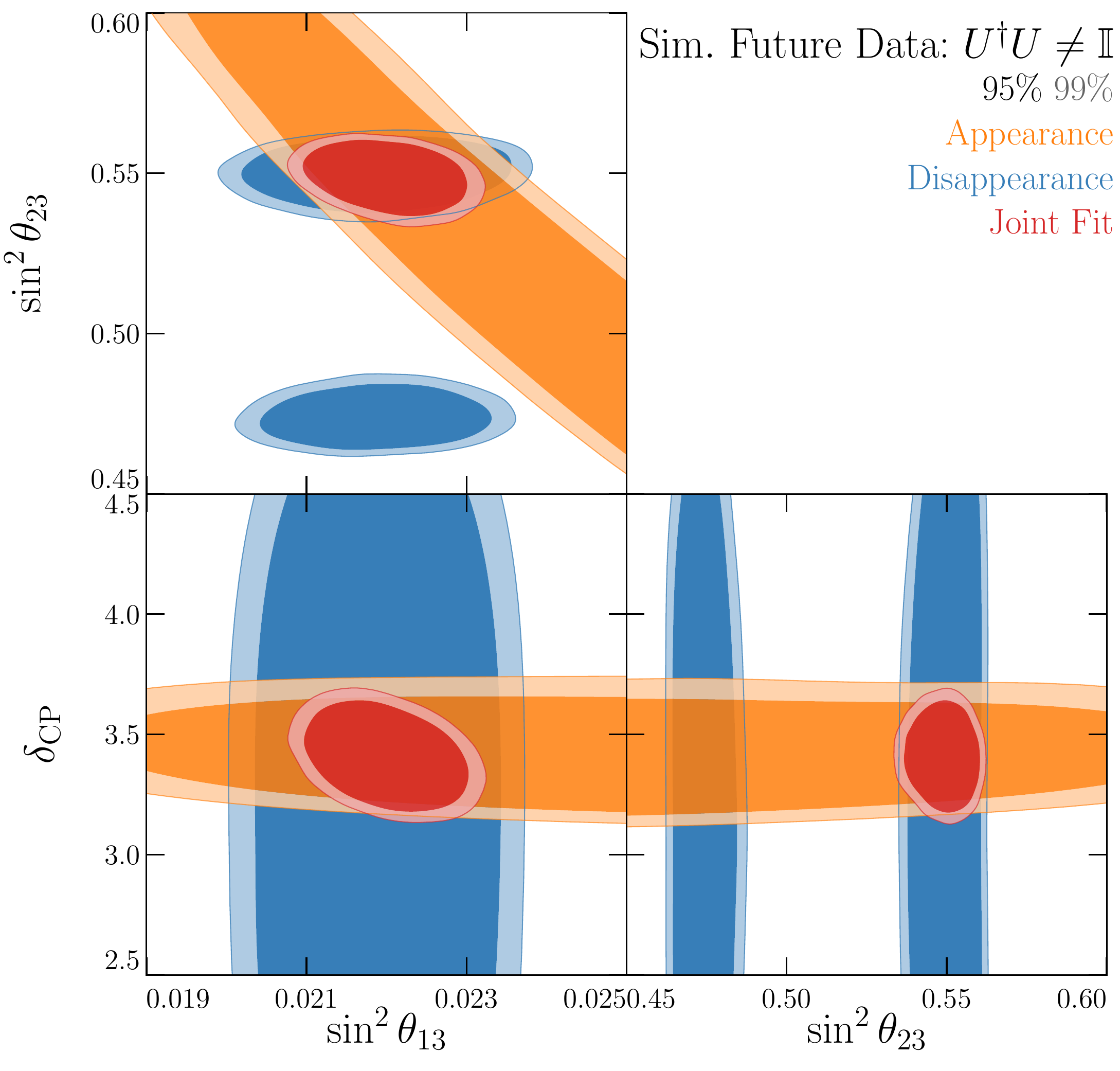}
\caption{Independent measurements of the parameters $\sin^2\theta_{13}$, $\sin^2\theta_{23}$, and $\delta_{\rm CP}$ by the appearance dataset (orange), disappearance dataset (blue), and their combination (red) when data are simulated with a non-unitary LMM, but analyzed assuming it is. The simulated data analyzed here are the same as those analyzed in Fig.~\ref{fig:unitarity_violation}. \label{fig:unitarity_violation_standardparams}}
\end{center}
\end{figure*}
%%%%%%%%%%%%%%%%%%%%%%%%%%%%%%%%%

From Eq.~\eqref{eq:rhoeta_emu}, it is apparent that some knowledge of $\delta_{\rm CP}$ is required in order to prefer a direction in the $(\rho_{e\mu}, \eta_{e\mu})$ plane, an effect seen in the blue regions throughout all six panels of Fig.~\ref{fig:UnTriangles}. Here, we see the combined measurement of disappearance channels, specifically driven by the future DUNE, T2HK, and JUNO experiments, choosing a preferred direction, i.e., being sensitive to the value of $\delta_{\rm CP}$. This is driven by the $\nu_\mu$/$\overline{\nu}_\mu$ disappearance measurements from DUNE/T2HK, which are sensitive to the quantities $\absq{U_{\mu 1}}$ and $\absq{U_{\mu 2}}$ independently at some level. With the assumed true values of the mixing angles and $\delta_{\rm CP}$, 
\begin{align}
\absq{U_{\mu1}} &= s_{12}^2 c_{23}^2 + c_{12}^2 s_{23}^2 s_{13}^2 + 2\cos\delta_{\rm CP} s_{12} c_{12} s_{13} s_{23} c_{23} \nonumber \\
&\approx 0.083, \\
\absq{U_{\mu2}} &= c_{12}^2 c_{23}^2 + s_{12}^2 s_{23}^2 s_{13}^2 - 2\cos\delta_{\rm CP} s_{12} c_{12} s_{23} s_{23} c_{23} \nonumber \\
&\approx 0.378.
\end{align}
For $\delta_{\rm CP} \approx 0$, $2\pi$, these values approach roughly $0.215$ and $0.247$, respectively. While this is a minor effect at DUNE and T2HK, their combination\footnote{Changing $\delta_{\rm CP}$ in this way leads to an effective ``tilt'' in the expected number of muon-neutrino events in the DUNE/T2HK disappearance channels, shifting the relative number of events below/above the disappearance minimum in the oscillation probability. While our simulations of T2HK and DUNE include normalization uncertainties, we do not include a spectral shape uncertainty which would mask this effect.} can distinguish between these $\absq{U_{\mu2}} > \absq{U_{\mu1}}$ and $\absq{U_{\mu1}} \approx \absq{U_{\mu2}}$ scenarios at ${\sim}99\%$ confidence, leading to the non-closure of our blue contours throughout Fig.~\ref{fig:UnTriangles}. However, we note here that this does \textit{not} imply that these experiments are sensitive to CP violation, as they are only measuring $\cos\delta_{\rm CP}$, a CP-symmetric quantity. Only with $\nu_e$ appearance (or related channels) can the amount of CP violation in the lepton sector, i.e., the Jarlskog invariant, be inferred. This is apparent by the reflective symmetry of each of the blue contours in Fig.~\ref{fig:UnTriangles} about the $\eta_{xy} = 0$ axis in each panel.

In order to test how unitarity violation would appear in these triangles, we simulate data with injected non-unitarity, but analyze it assuming the LMM is unitary. To include effects of non-unitarity, we adopt a parameterization beyond PMNS~\cite{FernandezMartinez:2007ms,Escrihuela:2015wra,Li:2015oal}, which requires 13 parameters, as described in detail in Appendix~\ref{sec:nonUparameterization}. We inject non-unitarity by making $\sum_i U_{ei} U_{\mu i}^* = 0.01 + 0.04i$ and $\sum_\alpha U_{\alpha 2} U_{\alpha 3}^* = -0.004+0.017i$, on the edge of what is allowed by current data.

Figure~\ref{fig:unitarity_violation} shows the constructed triangles in the $e$-$\mu$ and 2-3 planes, where the fit incorrectly assumes the LMM is unitary. With a joint disappearance and appearance analysis (red), there is no indication of unitarity violation as the joint contours appear similar in shape and size to those in Fig.~\ref{fig:UnTriangles}. However, individual channel measurements reveal tension: disappearance and appearance measurements disagree at over $95\%$ in the 2-3 plane. This demonstrates that separate analyses of disappearance and appearance measurements can be a powerful probe of unitarity violation in the lepton sector, and complementary to sterile neutrino searches~\cite{Athanassopoulos:1996wc,Athanassopoulos:1997pv,Armbruster:2002mp,Astier:2003gs,Astier:2001yj,Eskut:2007rn,Adamson:2017uda,Aguilar-Arevalo:2018gpe,Dentler:2018sju,Diaz:2019fwt,Boser:2019rta,Berryman:2019hme}. 

Importantly, no tension is present (even at $1\sigma$) when viewed in terms of measurements of the PMNS parameters even when disappearance and appearance channels are separate. We demonstrate this in Fig.~\ref{fig:unitarity_violation_standardparams}, where we show the same sets of experiments measuring this simulated data in terms of the parameters $\sin^2\theta_{13}$, $\sin^2\theta_{23}$, and $\delta_{\rm CP}$. Note that all three of these measurements intersect. This shows an important advantage of unitarity triangles as a test of LMM unitarity.

Testing whether certain subsets of data agree on these planes is one test of unitarity. However, constructing the triangles assumes the LMM is unitary. To numerically constrain non-unitarity, one needs to discard this assumption, and include sterile neutrino searches. This leads to non-intuitive connections between unitarity triangles and numerical results, e.g., the unitarity violation in $e$-$\mu$ plane for Fig.~\ref{fig:unitarity_violation} will be excluded by future searches at over $2\sigma$, contrary to what the figure suggests, when sterile search results are included. In what follows, we develop another intuitive means of using oscillation measurements to probe LMM unitarity, accounting for information loss when triangles are constructed under the assumption of unitarity. Specifically, we consider the consistency between measurements of the degree of CP violation in $U_{\rm LMM}$.

%%%%%%%%%%%%%%%%%%%%%%%%%%%%%%%%%%%%%%%%%%%%%%%%%%%%%%%%
%%%%%%%%%%%%%%%%%%%%%%%%%%%%%%%%%%%%%%%%%%%%%%%%%%%%%%%%

\section{Jarlskog Factors and the Jarlskog Invariant}
\label{sec:Jarlskog}

For a unitary LMM, the Jarlskog invariant [Eq.~(\ref{eq:Jpmns})] is a measure of CP violation and is related to the area of the unitarity triangles in Fig.~\ref{fig:UnTriangles}. By constructing the triangles as in Eq.~(\ref{eq:rhoetatext}), the area of each triangle is $(J_{\rm PMNS}/2)/(|U_{\alpha i}|^2 |U_{\beta j}|^2)$. As long as  $\delta_{\rm CP}$ is not $0$ or $\pi$, $J_{\rm PMNS}$, and therefore the triangle areas, are non-zero. In general, Jarlskog Factors $J_{\alpha i}$ can be calculated by taking the cofactor of the $(\alpha,i)$ element of the LMM, taking the complex conjugate of two elements. If the LMM is not unitary, 
these nine different cofactors need not be the same~\cite{Gandhi:2015xza}. 
 
The Jarlskog factors are defined as in Eq.~\eqref{eq:Jarlskog}, such that the nine $|J_{\alpha i}|$ are the same and equal to $|J_{\rm PMNS}|$ when the matrix is unitary. This condition is necessary, but not sufficient, for LMM unitarity. Without the unitarity assumption, six triangles provide information on at most six different $J_{\alpha i}$ (recall the discussion of Section~\ref{sec:LMM}). Therefore, to obtain a full characterization of the potentially non-unitary LMM, six unitarity triangles do not suffice. One solution is to construct nine triangles corresponding to the nine Jarlskog factors. However, as there are only six closure relations (Eq.~\ref{eq:closuretext}), this leads to some redundancy. Therefore, a compact manner of representing all six closure relations, as well as characterizing the full LMM is to show six unitarity triangles assuming unitarity, and nine Jarlskogs without assuming unitarity.

%%%%%%%%%%%%%%%%%%%%%%%%%%%%%%%%%
\begin{figure}[t]
\begin{center}
\includegraphics[width=0.8\columnwidth]{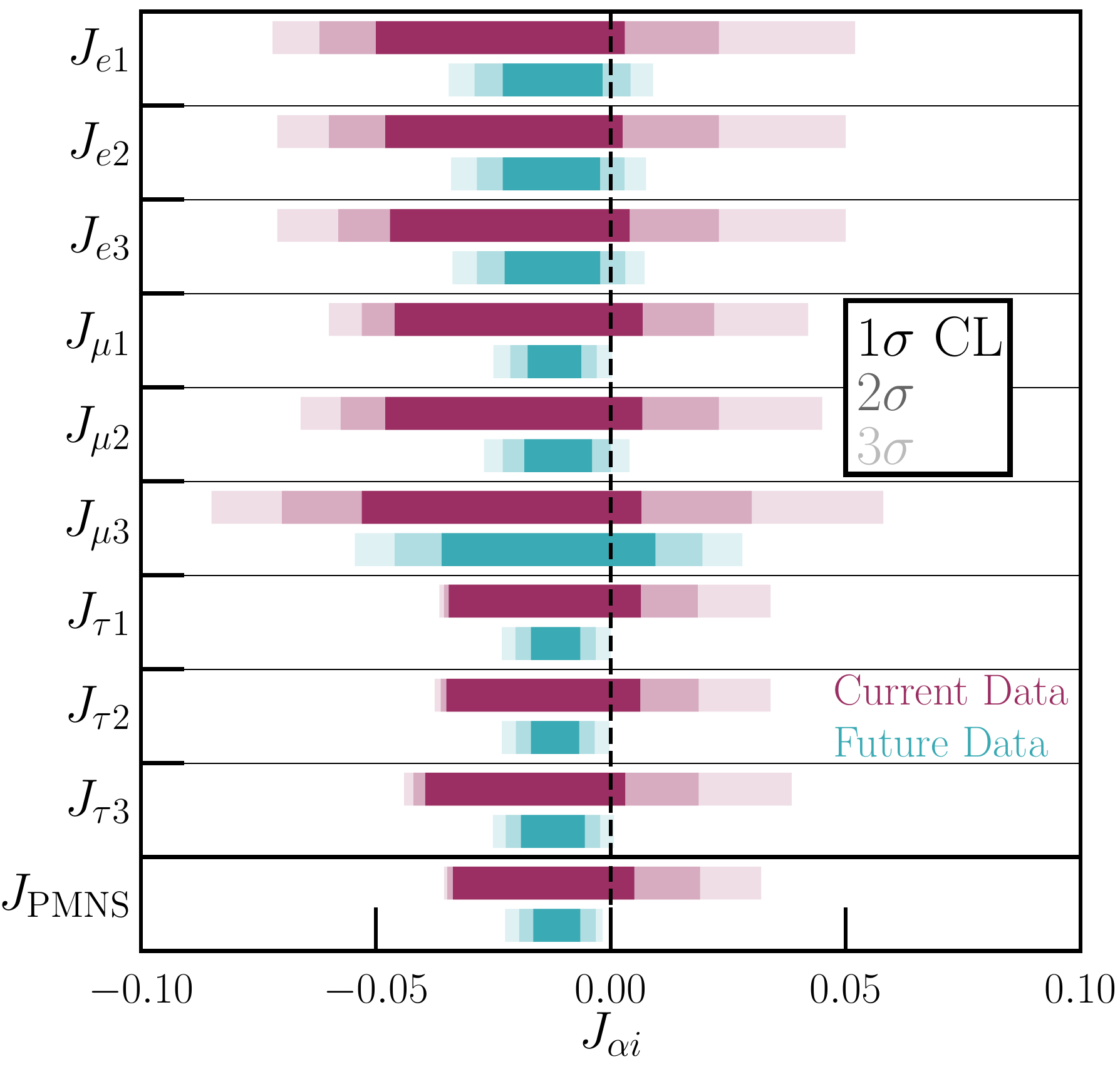}
\caption{Current (purple) and expected future (teal) 1, 2, and $3\sigma$ measurement ranges of the nine independent Jarlskog factors when not assuming unitarity, $J_{\alpha i}$, compared with the allowed range when assuming unitarity, $J_{\rm PMNS}$.  An inconsistency between the measured $J_{\alpha i}$ and $J_{\rm PMNS}$ would imply unitarity violation.
\label{fig:Jarlskogs}}
\end{center}
\end{figure}
%%%%%%%%%%%%%%%%%%%%%%%%%%%%%%%%%

To compare the Jarlskog invariant and Jarlskog factors, we perform a fit to all current and current plus future data while (not) assuming unitarity of the LMM to construct $J_{\rm PMNS}$ ($J_{\alpha i}$). When not assuming unitarity, we adopt the parameterization explained in Appendix~\ref{sec:nonUparameterization}. Figure~\ref{fig:Jarlskogs} shows the result. Our current measurement of $J_{\rm PMNS}$ (bottom row, purple) is consistent with the results of other more detailed fits~\cite{Esteban:2020cvm}. Each independent $J_{\alpha i}$ measurement agrees, consistent with the unitary LMM hypothesis.  We see that $J_{\rm PMNS}$ and $J_{\alpha i}$ are all consistent with zero at 1$\sigma$, consistent with the hypothesis that CP is conserved in the lepton sector. When we simulate future data (assuming $J_{\alpha i}$ are all equal and nonzero), the inclusion of future JUNO, IceCube, DUNE, and T2HK projections allows us to exclude $J_{\rm PMNS} = 0$ at $3\sigma$. Moreover, many of the different $J_{\alpha i}$ will disfavor CP-conservation at $3\sigma$. However, certain Jarlskog Factors, particularly those involving knowledge of $|U_{\tau i}|$ (especially $J_{\mu 3}$), will remain difficult to measure when unitarity is not assumed.

%%%%%%%%%%%%%%%%%%%%%%%%%%%%%%%%%%%%%%%%%%%%%%%%%%%%%%%%
%%%%%%%%%%%%%%%%%%%%%%%%%%%%%%%%%%%%%%%%%%%%%%%%%%%%%%%%

\section{Conclusions}
\label{sec:Conclusions}

We have presented a comprehensive analysis of leptonic unitarity triangles using all current and future neutrino oscillation data. Figure~\ref{fig:UnTriangles} displays how the closure of six unitarity triangles is/will be constrained. By virtue of the nature of these measurements, in contrast with the CKM matrix, intersections of many measurements of PMNS matrix parameters are not possible. Non-unitarity can nevertheless explicitly manifest itself as shown in Fig.~\ref{fig:unitarity_violation}, though observation of non-unitarity requires distinguishing between appearance and disappearance datasets. Figure~\ref{fig:Jarlskogs} presents an alternative and complementary visualization of constraints on LMM unitarity in terms of Jarlskog factors. The allowed ranges of the $J_{\alpha i}$ are consistent with each other and with non-zero CP violation in the lepton sector. If the LMM was not unitary, these measurements would yield different $J_{\alpha i}$.

The Standard Model demands new physics to explain the origin of neutrino masses and therefore oscillations. The impending era of precision experiments will enable us to further understand the structure of the leptonic mixing matrix, and constraints on the matrix's unitarity serve as a probe of the mechanism of neutrino masses. Meanwhile, the origin of the baryon asymmetry of the universe will be better understood through studies of the degree of CP violation in the lepton sector. Performing detailed studies of the leptonic unitarity triangles therefore bears directly on both of these problems in the Standard Model.

%%%%%%%%%%%%%%%%%%%%%%%%%%%%%%%%%%%%%%%%%%%%%%%%%%%%%%%%
%%%%%%%%%%%%%%%%%%%%%%%%%%%%%%%%%%%%%%%%%%%%%%%%%%%%%%%%

\vspace{0.7cm}
\centerline{\bf Acknowledgments}
\vspace{0.2cm}
For helpful discussions, we are grateful to
Francesco Capozzi,
Andr\'e de Gouv\^ea,
Peter Denton,
Pedro Machado,
Stephen Parke,
Michael Peskin,
Xin Qian, 
and Natalia Toro.
We acknowledge the 7th LCTP Spring Symposium: Neutrino Physics. SARE and SWL are supported by the U.S. Department of Energy under Contract No. DE-AC02-76SF00515. SARE is also supported in part by the Swiss National Science Foundation, SNF project number P400P2$\_$186678. KJK is supported by Fermi Research Alliance, LLC under contract DE-AC02-07CH11359 with the U.S. Department of Energy.

%%%%%%%%%%%%%%%%%%%%%%%%%%%%%%%%%%%%%%%%%%%%%%%%%%%%%%%%
%%%%%%%%%%%%%%%%%%%%%%%%%%%%%%%%%%%%%%%%%%%%%%%%%%%%%%%%

% \bibliographystyle{apsrev4-1}
% \bibliography{refs}

%%%%%%%%%%%%%%%%%%%%%%%%%%%%%%%%%%%%%%%%%%%%%%%%%%%%%%%%
%%%%%%%%%%%%%%%%%%%%%%%%%%%%%%%%%%%%%%%%%%%%%%%%%%%%%%%%

\begin{appendix}

%%%%%%%%%%%%%%%%%%%%%%%%%%%%%%%%%%%%%%%%%%%%%%%%%%%%%%%%
%%%%%%%%%%%%%%%%%%%%%%%%%%%%%%%%%%%%%%%%%%%%%%%%%%%%%%%%

\setcounter{figure}{0}
\renewcommand{\thefigure}{A\arabic{figure}}
\renewcommand{\theHfigure}{A\arabic{figure}}

\titleformat{\section}[hang]{\bfseries\center\small}{}{1em}{Appendix \thesection.\quad#1}%
\titleformat{\subsection}[hang]{\bfseries\small\center}{}{1em}{\thesection.\thesubsection\quad#1}%

% \widetext
% \begin{center}
%  \bf \large Supplemental Material
% \end{center}
% \vspace*{0.2cm}

%%%%%%%%%%%%%%%%%%%%%%%%%%%%%%%%%%%%%%%%%%%%%%%%%%%%%%%%
%%%%%%%%%%%%%%%%%%%%%%%%%%%%%%%%%%%%%%%%%%%%%%%%%%%%%%%%

\section{Parameterization of the LMM}
\label{sec:rhoeta}

The unitarity conditions for the 3$\times$3 LMM are
\begin{equation}
    U^\dagger U = U U^\dagger = \mathbb{I} \ .
\end{equation}
From this, one can write down six real equations, requiring the normalization of rows and columns of matrix-elements-squared $|U_{\alpha i}|^2$ to be 1:
\begin{equation}\label{eq:Closures}
    \begin{array}{c c}\displaystyle\sum_i |U_{\alpha i}|^2 = 0\ ,\\ \text{(rows)}\end{array} \quad  \begin{array}{cc}
   \displaystyle\sum_\alpha |U_{\alpha i}|^2 = 0 \ . \\ \text{(columns)} \end{array} 
\end{equation}

One can also write six complex equations corresponding to ``closures'' between two different rows ($\alpha$ and $\beta$) or two different columns ($i$ and $j$):
\begin{align}
    \sum_{i=1}^{3} U_{\alpha i} U_{\beta i}^* = \, &U_{\alpha 1} U_{\beta 1}^* + U_{\alpha 2} U_{\beta 2}^* + U_{\alpha 3} U_{\beta 3}^* = 0,\quad \alpha\neq\beta \nonumber\\
    &\text{(row closure)}, \\
    \sum_{\alpha = e}^{\tau} U_{\alpha i} U_{\alpha j}^* = \, &U_{e i} U_{ej}^* + U_{\mu i} U_{\mu j}^* + U_{\tau i} U_{\tau j}^* = 0,\quad i\neq j \nonumber \\
    &\text{(column closure)}.
\end{align}
From these closure relations, one can construct the familiar unitarity triangles in the $(\rho,\eta)$ plane by dividing each term in the closure by one of the sides. For a given row/column, there are 3 different triangles one could define that are not related to one another by a simple inversion. The triangles are defined as in the main text in terms of the LMM matrix elements.

\begin{widetext}
The chosen set of $(\rho,\eta)$ used in the main text, under the assumption that the LMM is unitary, can be expressed as:~\footnote{
\vspace{-2em}
\begin{minipage}{1.0\textwidth}
A unitary LMM can be written in the usual form in terms of the PMNS rotation angles and CP-violating phase as:
\end{minipage}
\vspace{2em}
\begin{equation*}
    \quad\
    U_{\rm LMM} = \Unu = \begin{pmatrix} 
    c_{12} c_{13} & s_{12} c_{13} & s_{13} e^{-i \delta_{\rm CP}} \\
    -s_{12}c_{23} - c_{12}s_{13}s_{23}e^{i \delta_{\rm CP}} & c_{12}c_{23} - s_{12}s_{13}s_{23}e^{i \delta_{\rm CP}} & c_{13} s_{23} \\
    s_{12}s_{23} - c_{12}s_{13}c_{23}e^{i \delta_{\rm CP}} & -c_{12}s_{23} - s_{12}s_{13}c_{23}e^{i \delta_{\rm CP}} & c_{13} c_{23}
    \end{pmatrix} \ .
\end{equation*}
}
\begin{align}
\nonumber &\rho_{e\mu} = c_{12}^2 + \cos\delta_{\rm CP} \left( \frac{s_{12} c_{12}}{s_{13} t_{23}}\right) ,
\\
\nonumber
\\
&\eta_{e\mu} = \sin\delta_{\rm CP} \left( \frac{s_{12} c_{12}}{s_{13} t_{23}}\right) ,
\\
\nonumber
\\
%\end{align}
%\begin{align}
\nonumber &\rho_{e\tau} = \frac{1}{2}\left(\frac{
2 s_{12}^2( s_{23}^2 -c_{23}^2 s_{13}^2) - t_{12} s_{13} \sin2\theta_{23} \cos2\theta_{12} \cos \delta_{\rm CP} }
{c_{12}^2 s_{13}^2 c_{23}^2 +s_{12}^2 s_{23}^2-2\Delta \cos\delta_{\rm CP}}\right)\ , 
\\
&\eta_{e\tau} = -\frac{1}{2}\left(\frac{t_{12}s_{13} \sin2\theta_{23} \sin\delta_{\rm CP}}
{c_{12}^2 s_{13}^2 c_{23}^2 +s_{12}^2 s_{23}^2-2\Delta \cos\delta_{\rm CP}}\right) \ , 
\\
\nonumber
\\
%\end{align}
%\begin{align}
\nonumber &\rho_{\mu \tau}
= \frac{c_{13}^2}{4}\left(\frac{\sin^2 2\theta_{23}\left(c_{12}^2 - s_{12}^2 s_{13}^2\right) +4\cos2\theta_{23}\Delta\cos \delta_{\rm CP}}
{\left(c_{12}^2 c_{23}^2 + s_{12}^2 s_{13}^2 s_{23}^2 -2\Delta \cos\delta_{\rm CP}\right)\left(c_{12}^2 s_{23}^2 + s_{12}^2 s_{13}^2 c_{23}^2 + 2\Delta \cos\delta_{\rm CP}\right)}\right) \ ,
\\
&\eta_{\mu \tau}  
= c_{13}^2\frac{\Delta\sin \delta_{\rm CP}}
{\left(c_{12}^2 c_{23}^2 + s_{12}^2 s_{13}^2 s_{23}^2 -2\Delta \cos\delta_{\rm CP}\right)\left(c_{12}^2 s_{23}^2 + s_{12}^2 s_{13}^2 c_{23}^2 + 2\Delta \cos\delta_{\rm CP}\right)} \ ,
%\\
%\nonumber
%\\
\end{align}
\begin{align}
\nonumber &\rho_{12}
= \frac{c_{13}^2}{4}\left(\frac{\sin^22\theta_{12}\left(c_{23}^2 - s_{13}^2 s_{23}^2\right) + 4\cos2\theta_{12}\Delta \cos \delta_{\rm CP}}
{\left(s_{12}^2 c_{23}^2 + c_{12}^2 s_{13}^2 s_{23}^2 +2\Delta \cos\delta_{\rm CP}\right)\left(c_{12}^2 c_{23}^2 + s_{12}^2 s_{13}^2 s_{23}^2 - 2\Delta \cos\delta_{\rm CP}\right)}\right) \ ,
\\
&\eta_{12}  
= -c_{13}^2\frac{\Delta \sin \delta_{\rm CP}}
{\left(s_{12}^2 c_{23}^2 + c_{12}^2 s_{13}^2 s_{23}^2 +2\Delta \cos\delta_{\rm CP}\right)\left(c_{12}^2 c_{23}^2 + s_{12}^2 s_{13}^2 s_{23}^2 - 2\Delta \cos\delta_{\rm CP}\right)} \ ,
\\
\nonumber
\\
\nonumber &\rho_{13}
= \frac{1}{2}\left(\frac{2s_{23}^2\left(s_{12}^2 - s_{13}^2 c_{12}^2\right) - t_{23} s_{13} \sin2\theta_{12} \cos2\theta_{23}\cos \delta_{\rm CP}}
{s_{12}^2 s_{23}^2 + c_{12}^2 s_{13}^2 c_{23}^2 - 2\Delta \cos\delta_{\rm CP}}\right) \ ,
\\
&\eta_{13} 
= \frac{1}{2}\left(\frac{t_{23} s_{13} \sin2\theta_{12} \sin \delta_{\rm CP}}
{s_{12}^2 s_{23}^2 + c_{12}^2 s_{13}^2 c_{23}^2 - 2\Delta \cos\delta_{\rm CP}}\right) \ ,
\\
\nonumber
\\
\nonumber &\rho_{23}
= c_{23}^2\left(1+\frac{t_{23}\cos \delta_{\rm CP}}
{t_{12}s_{13}}\right) \ ,
\\
&\eta_{23}  
= -c_{23}^2\left(\frac{t_{23} \sin \delta_{\rm CP}}
{t_{12} s_{13}}\right) \ ,
\end{align}
where $4\Delta \equiv s_{13}\sin2\theta_{12} \sin2\theta_{23}$.
\end{widetext}

%%%%%%%%%%%%%%%%%%%%%%%%%%%%%%%%%%%%%%%%%%%%%%%%%%%%%%%%
%%%%%%%%%%%%%%%%%%%%%%%%%%%%%%%%%%%%%%%%%%%%%%%%%%%%%%%%

\section{Non-Unitarity Parameterization \& Effects of Assuming Unitarity}
\label{sec:nonUparameterization}

In this appendix we give more detail regarding the $U_{\rm LMM}$ parameterization that is used when unitarity is not manifestly assumed (as in the PMNS parameterization). We also show how such assumptions impact the measurements shown in unitarity triangles like those in Fig.~\ref{fig:UnTriangles}.

When we do not assume that the LMM is unitarity, we assume that the it takes the form
\begin{equation}
    U_{\rm LMM} = \left(\begin{array}{l l l}
    \left\lvert U_{e1}\right\rvert & \left\lvert U_{e2} \right\rvert e^{i\phi_{e2}} & \left\lvert U_{e3}\right\rvert e^{i\phi_{e3}} \\
    \left\lvert U_{\mu1}\right\rvert & \left\lvert U_{\mu2} \right\rvert & \left\lvert U_{\mu3}\right\rvert \\
    \left\lvert U_{\tau1}\right\rvert & \left\lvert U_{\tau2} \right\rvert e^{i\phi_{\tau2}} & \left\lvert U_{\tau3}\right\rvert e^{i\phi_{\tau3}} \\
    \end{array}\right)\ ,
\end{equation}
where the 13 free parameters (ignoring the potentially physical Majorana phases) are necessary to describe a $3\times 3$ complex matrix after accounting for rephasing of the charged lepton fields. Alternative parameterizations are commonly adopted in the literature~\cite{FernandezMartinez:2007ms,Escrihuela:2015wra,Li:2015oal}, all with the same number of free parameters. We motivate the use of our parameterization, and discuss maps between this and others in the literature in Ref.~\cite{Ellis:2020hus}. While the complex phases in $U_{\rm LMM}$ appear on different matrix elements than in $\Unu$, the two parameterizations are related (if $U_{\rm LMM}$ satisfies the unitarity conditions) by rephasing of the neutrino fields. For any set of PMNS parameters, an equivalent set of LMM parameters may be determined uniquely.

We obtain best-fit values for the 13 LMM parameters by making use of the observation that the LMM fit must match the global fit for the PMNS parameters when the LMM is unitary. Analyzing current data when assuming unitarity yields the maximum-likelihood values of $\sin^2\theta_{12} = 0.308$, $\sin^2\theta_{13} = 0.0219$, $\sin^2\theta_{23} = 0.551$, and $\delta_{\rm CP} = 200^\circ$. These four best-fit values may then be used in conjunction with the 9 constraints applicable to a $3\times3$ unitary matrix to solve for the 13 LMM elements, yielding
\begin{align}\label{eq:LMMBestFit}
    &\left\lvert U_{\rm LMM} \right\rvert = \begin{pmatrix}
    0.823 & 0.549 & 0.148 \\
    0.288 & 0.615 & 0.734 \\
    0.490 & 0.555 & 0.663 \\
    \end{pmatrix}\ ,\nonumber\\ 
    &\begin{array}{cc}
    \phi_{e2} = 172^\circ, & \phi_{e3} = 333^\circ, \\ \phi_{\tau2} = 346^\circ, & \phi_{\tau 3} = 170^\circ .
    \end{array}
\end{align}
When we study the nine separate Jarlskog factors $J_{\alpha i}$ cf. Fig.~\ref{fig:Jarlskogs}, we use this set of 13 parameters, projecting down to relevant combinations for the allowed regions of different $J_{\alpha i}$. 

We also use this parameterization to simulate future data for JUNO IceCube, T2HK, and DUNE assuming $U_{\rm LMM}$ is \textit{not} unitary, i.e., $U^\dagger U \neq \mathbb{I}$. In generating the data that are analyzed for Fig.~\ref{fig:unitarity_violation}, we modify the input values of $\phi_{e2}$ and $\phi_{e3}$ from those in Eq.~(\ref{eq:LMMBestFit}) by $\Delta \phi_{e2} = -5.30^\circ$ and $\Delta \phi_{e3} = 7.23^\circ$. This preserves the unitarity constraint that the individual rows and columns of $U_{\rm LMM}$ are properly normalized, $\sum_i |U_{\alpha i}|^2 = 1$, $\sum_\alpha |U_{\alpha i}|^2 = 1$, but causes non-closure of the triangles in several planes.

%%%%%%%%%%%%%%%%%%%%%%%%%%%%%%%%%
\begin{figure*}[t]
    \centering
    \includegraphics[width=0.8\textwidth]{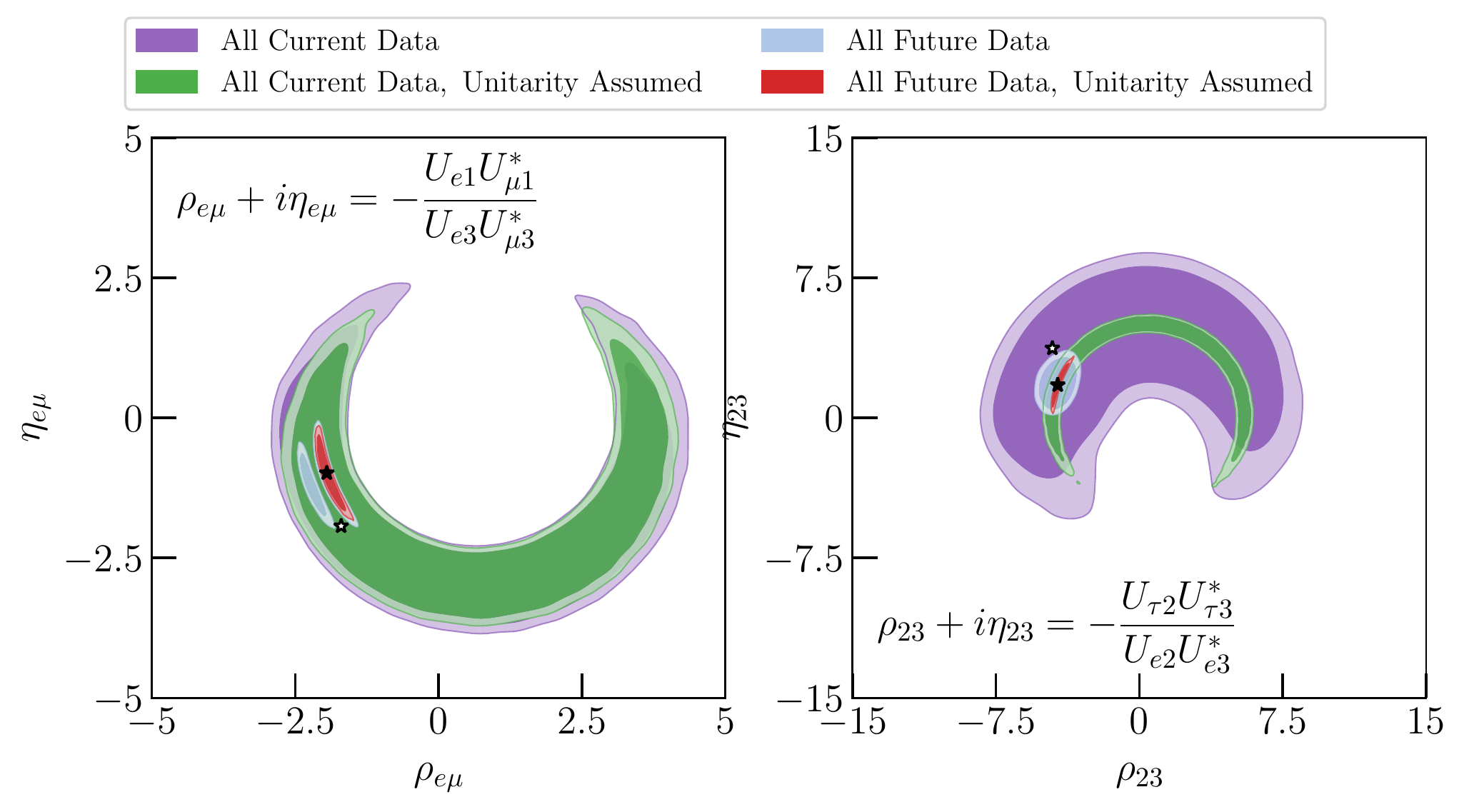}
    \caption{Current (purple and green) and expected future (pale blue and red) measurements  95\% (dark colors) and 99\% confidence level (light) of two different unitarity triangles -- $\rho_{e\mu}$ vs. $\eta_{e\mu}$ (left) and $\rho_{23}$ vs. $\eta_{23}$ (right). We contrast two assumptions in this figure, showing the resulting measurements when the unitarity of the leptonic mixing matrix is or is not assumed. Purple and light blue contours display the results when unitarity is not assumed, where green and red contours show the results when it is assumed. The filled-in (open) star indicates the best-fit point of the analysis of current data when unitarity is (not) assumed, corresponding to the green (purple) contours.}
    \label{fig:RhoEta_UnitarityAssumptions}
\end{figure*}
%%%%%%%%%%%%%%%%%%%%%%%%%%%%%%%%%

One non-closure is in the $e$-$\mu$ plane, with $\sum_i U_{ei} U_{\mu i}^* = 0.01 + 0.04i$. Sterile neutrino searches that look for zero-distance neutrino oscillation (as described in Section~\ref{app:CurrentData}) are sensitive to the absolute value squared of the non-closure, and this level is at the upper end of what is currently allowed by data. In addition, there is also non-closure in the 2-3 plane, with $\sum_\alpha U_{\alpha 2} U_{\alpha 3}^* = -0.004+0.017i$.  This is shown in Fig.~\ref{fig:unitarity_violation}.

\textbf{Triangles when Unitarity is not assumed:} In the results shown in Fig.~\ref{fig:UnTriangles} and Fig.~\ref{fig:unitarity_violation}, fits to existing/future data were performed with the PMNS mixing angles as free parameters ($\sin^2\theta_{12}$, $\sin^2\theta_{13}$, $\sin^2\theta_{23}$, $\delta_{\rm CP}$, and mass-squared splittings), such that unitarity was explicitly assumed. Confidence level contours were then constructed by mapping these parameters onto those that enter each unitarity triangle, using the PMNS parameterization. Here we perform a fit using the parameterization described above, where the unitarity of $U_{\rm LMM}$ is not guaranteed.

The difference between these two fits is most apparent in triangles that depend on mixing matrix elements that are not powerfully measured on their own in experiments, but can be inferred by other measurements if unitarity is assumed. Specifically, that is the case for the mixing matrix elements $U_{\tau i}$. For instance, in the PMNS parameterization, $U_{\tau 3} = \cos{\theta_{23}} \cos{\theta_{13}}$, which can be constrained fairly well by atmospheric and reactor experiments. Without unitarity, the strongest current measurement of $U_{\tau 3}$ in our fit comes from OPERA's $\nu_\mu \to \nu_\tau$ appearance, a significantly less precise measurement.

We show this difference in Fig.~\ref{fig:RhoEta_UnitarityAssumptions}, focusing on two different triangle planes, ($\rho_{e\mu}$, $\eta_{e\mu}$), where the differences are small, and ($\rho_{23}$, $\eta_{23}$), where the differences are the largest. All contours shown are 95\% (dark contours) and 99\% confidence level (light). Here, the purple and green regions correspond to current data fits, where the purple (green) region is the result of the fit without (with) assuming unitarity. Likewise, light blue (unitarity not assumed) and red (unitarity assumed) regions are from fits including current and future data. The green and red regions here correspond with those of the same color shown in the appropriate panels of Fig.~\ref{fig:UnTriangles}. In each of the two panels, the filled-in star denotes the best-fit point in this parameter space obtained by analyzing all current data when unitarity is assumed (green dataset), where the open star indicates the best fit point when unitarity is not assumed (purple dataset).

Most notable here is the difference in the size of allowed regions between when unitarity is or is not assumed for ($\rho_{23}$, $\eta_{23}$). As mentioned above, this is largely due to the uncertainty regarding the magnitude of the elements $|U_{\tau 2}|$ and $|U_{\tau 3}|$. We also see that the current data prefer a much larger triangle in this plane if unitarity is not assumed -- this is due to the preference for large $|U_{\tau 3}|$ from the OPERA experiment~\cite{Acero:2019ksn}. We also highlight the second island of allowed parameter space in the future projections of ($\rho_{e\mu}$, $\eta_{e\mu}$) when unitarity is not assumed (light blue) -- this comes about because, when unitarity is not assumed, future experiments cannot definitively determine whether $\left\lvert U_{\mu 1}\right\rvert^2$ is smaller or larger than $\left\lvert U_{\mu 2}\right\rvert^2$.

%%%%%%%%%%%%%%%%%%%%%%%%%%%%%%%%%%%%%%%%%%%%%%%%%%%%%%%%
%%%%%%%%%%%%%%%%%%%%%%%%%%%%%%%%%%%%%%%%%%%%%%%%%%%%%%%%

\end{appendix}

\bibliographystyle{apsrev4-1}
\bibliography{refs}

\end{document}